\documentclass[9pt,a4paper,twocolumn]{article}
\pdfoutput=1
\usepackage{times}
\usepackage{graphicx}
\usepackage{dcolumn}
\usepackage{bm}

\usepackage[utf8]{inputenc}
\usepackage[T1]{fontenc}
\usepackage{mathptmx}
\usepackage{multirow}
\usepackage{xcolor}
\usepackage{footnote}
\usepackage{braket}
\usepackage{siunitx}
\usepackage{hyperref}
\usepackage{booktabs}
\usepackage{amsmath}
\usepackage[numbers]{natbib}
\usepackage{mhchem}

\begin{document}
{\bf \Large \noindent Towards detection of molecular
parity violation by microwave spectroscopy of CpRe(CH$_{3}$)(CO)(NO) $^{\ast\ast}$}

\bigskip

{\noindent \it \large Nityananda Sahu,$^{1}$ Konstantin Gaul,$^{1}$ Anke Wilm,$^{1}$, Melanie Schnell$^{2}$ and Robert Berger$^{1,\ast}$}

\bigskip

{\footnotesize 
\fontfamily{phv}\selectfont
\noindent $\left[ ^{1} \right]$ Dr. Nityananda Sahu, Dr. Konstantin
Gaul, Dr. Anke Wilm, Prof. Dr. Robert Berger \\
Fachbereich Chemie, Theoretische Chemie \\
Philipps-Universit{\"{a}}t Marburg, Hans-Meerwein-Str. 4, 35032 Marburg, Germany \\
E-mail: robert.berger@uni-marburg.de   \\
\noindent $\left[ ^{2} \right]$ Prof. Dr. Melanie Schnell \\
Deutsches Elektronen-Synchrotron DESY \\
Notkestr. 85, 22607 Hamburg, Germany \\
E-mail: melanie.schnell@desy.de   \\
$\left[ ^{\ast \ast} \right]$
This work is financially supported by the Deutsche Forschungsgemeinschaft (DFG, German
Research Foundation) -- Projektnummer 328961117 -- SFB 1319 ELCH. The center for scientific
computing (CSC) Frankfurt is thanked for computer time. \\
}

\today     \\ \\ 

\emph{Parity-violating differences in rotational constants of a chiral 5d
transition metal complex, that was previously experimentally well-characterised 
by broad-band microwave spectroscopy, are predicted with a recently
established efficient
analytical derivative technique. Relative differences $\Delta X/X$
between rotational constants $X=A,B,C$ of enantiomers of the title compound are
found to be on the order of $10^{-14}$, which is a favourably large effect.
The quality of the theoretical estimates is carefully assessed by computing
nuclear electric quadrupole coupling constants that agree well with 
experiment.}

\bigskip 

\begin{figure}[ht]
\centering
\includegraphics[width=\linewidth]{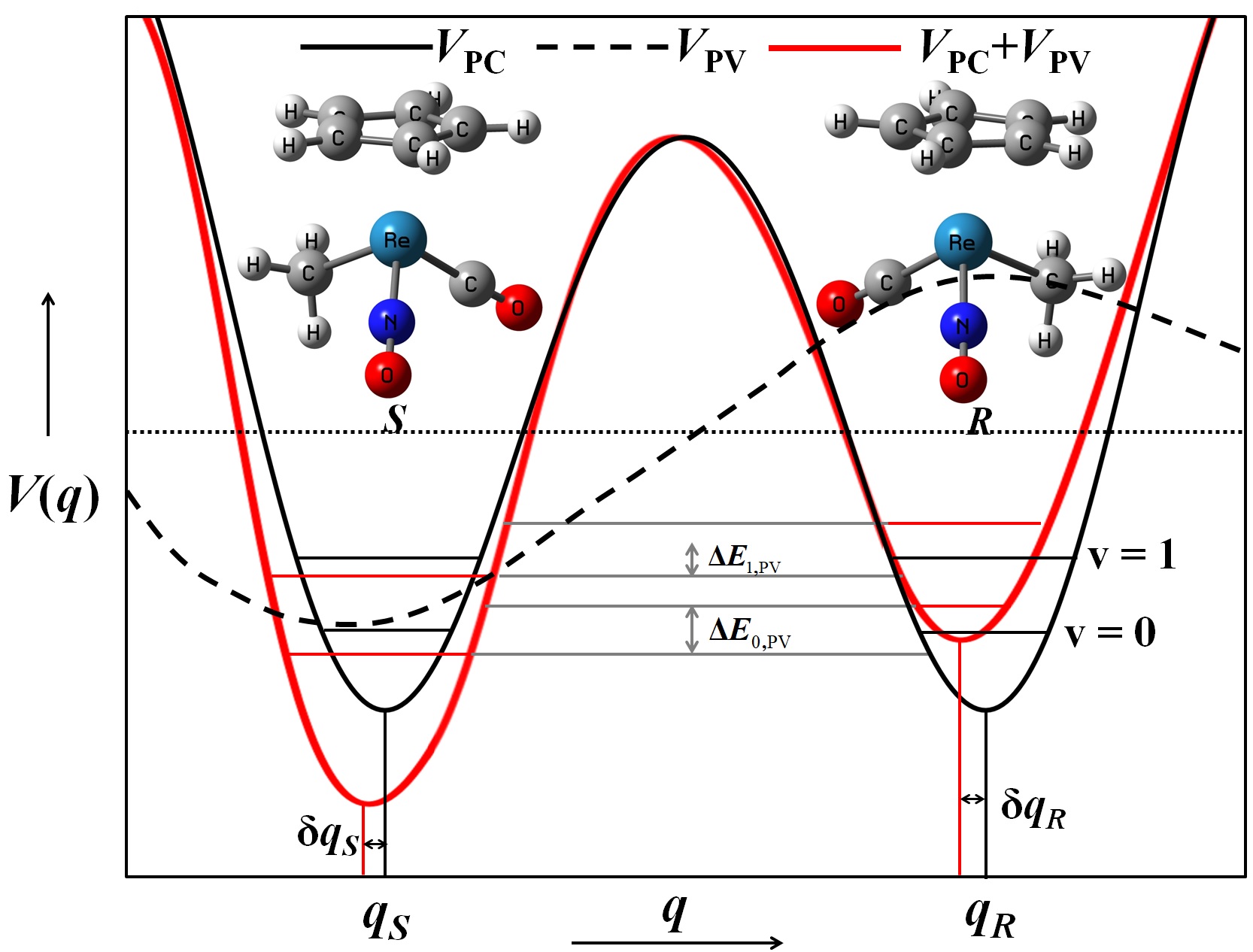}
\caption{Parity conserving (pc) and parity-violating (pv) energy profile diagram for {\it S}- and {\it R}-enantiomers of the CpRe(CH$_{3}$)(CO)(NO) complex. Energies are not to scale and PV contributions are magnified by several orders of magnitude.}
\label{fig:Re}
\end{figure}

One of the most intriguing effects of the fundamental weak force is
that it can induce a tiny energy difference ($\Delta E_\mathrm{pv}$)
between enantiomers of a chiral compound \cite{berger:2019}. Although
its existence has been predicted more than half a century ago
\cite{yamagata:1966,letokhov:1975,zeldovich:1977}, the
parity-violating energy difference escaped all attempted measurements,
despite its fundamental importance for our understanding of molecular
chirality, its potential role in the evolution of biomolecular
homochirality on planet earth \cite{quack:1989,quack:2002,berger:2019}
and its use to probe specific candidates of dark matter 
\cite{gaul:2020c,gaul:2020d}. Early experimental approaches
focused on vibrational spectroscopy, attempting to detect differences
in the vibrational transition frequencies of enantiomers (see Fig.~\ref{fig:Re})
\cite{letokhov:1975,kompanets:1976,arimondo:1977,bauder:1997,hollenstein:1997}.

The currently tightest upper bound from experiment, provided for the splitting 
in the C--F stretch fundamental in CHBrClF, \emph{the} prototypical chiral 
molecule \cite{costante:1997,pitzer:2013,pitzer:2015}, is on the order of
$\Delta\nu/\nu \approx 10^{-13}$ \cite{daussy:1999,ziskind:2002}. This
exceeds, however, the theoretically predicted splittings by about
four orders of magnitude
\cite{quack:2000,laerdahl:2000a,viglione:2000,quack:2003,berger:2007,bruck:2021,rauhut:2021}.
Compounds with heavier elements can, thus, become advantageous in this
research line, because parity violating energy differences coarsely scale
quintically with the nuclear charge $Z$ of the heaviest elements. The
single-center theorem \cite{hegstrom:1980} indicates that parity violating
effects independent of the nuclear spin are suppressed in main group
compounds with only a single heavy center, so that chiral methane
derivatives with two heavy halogens like CHAtFI were proposed as a remedy
\cite{berger:2007}. Although such short-lived radioactive molecules were
initially understood merely as theoretical toy models to explore limits in
high-$Z$ regions, recent progress based on our theoretical laser-cooling
proposal to explore parity violation \emph{within the nucleus} of various
isotopes of diatomic RaF \cite{isaev:2010} combined with the
subsequent first laser-spectroscopic investigation of this short-lived
molecule \cite{garciaruiz:2020} changed the picture completely and
initiated campaigns for molecular spectroscopy with radioisotopes at
high-energy physics centers such as CERN near Geneva, FRIB in Michigan or
TRIUMF in Vancouver to name but a few. 

Another way to elegantly bypass the
single-center theorem in chiral molecules is to turn to transition metal
compounds, as their additional d-orbital involvement can produce comparatively large
parity violating effects even in the absence of other heavy nuclei.
Schwerdtfeger and coworkers thus searched for known chiral Re and Ir
complexes and predicted their parity violating energy differences at the
equilibrium structure of the electronic ground state to be comparatively
large, on the order of 10 to 100 nJ~mol$^{-1}$ (or
10$^{-14}$~$E_\mathrm{h}$) \cite{schwerdtfeger:2003}. By virtue of
Lethokov's rule that relative parity violating energy splittings remain
essentially the same in electronic, vibrational and rotational spectra
\cite{letokhov:1975}, a Re complex and an Os complex with a double bond to
O and C, respectively, to facilitate strong dipole transitions in the
range of highly stable CO$_2$ lasers, were explored theoretically for large parity
violating vibrational frequency splittings. And indeed, exceptionally large
splittings of $\Delta\nu/\nu = 10^{-14}$ were finally estimated 
\cite{schwerdtfeger:2004b}. 

While highly encouraging as a result, two obstacles severely hampered
further progress for such complexes: 1) Ways to obtain enantiomerically
enriched samples of the chiral molecules remained open and 2) nothing was
known about the high-resolution gas phase spectroscopy of chiral transition
metal complexes. But advances in microwave spectroscopy provided the
solution to both: 1) Hirota's microwave three-wave mixing proposal to
selectively tag enantiomers of chiral molecules with C$_1$ symmetry
\cite{hirota:2012} was successfully implemented experimentally
\cite{patterson:2013} and makes spatial separation of enantiomers
unnecessary. 2) Medcraft et al.  reported the first high-resolution
spectroscopy of racemic \ce{CpRe(CH_3)(CO)(NO)} and analysed the
hyperfine-resolved rotational spectrum to obtain nuclear electric
quadrupole coupling (NEQC) constants \cite{medcraft:2014}. It was proposed
before \cite{schnell:2011} that precision microwave spectroscopy of slowed
chiral molecules might also be a promising route to detect parity-violating
frequency splittings, although up to this date little is known about the
magnitude of the corresponding line splittings in general and for
transition metal compounds in particular. The reason is that
parity-violation induced shifts in rotational constants in larger molecules
are extremely tedious to predict as this requires, even in the most simple
schemes, to compute gradients of parity violating potentials with respect
to spatial displacements of all the nuclei in the system. Thus, the rare
predictions attempted so far concern light molecules with few atoms only
like CHBrClF ($\Delta X/X \approx 10^{-17}$) \cite{quack:2000a} or fluorooxirane ($\Delta X/X \approx 10^{-19}$) \cite{berger:2001}, for
which these shifts were obtained with finite difference schemes in a
one-component, non-relativistic framework \cite{berger:2000a}. Our most
recent advances in theoretical approaches, however, allow such predictions
for polyatomic heavy-elemental compounds via an efficient analytic gradient
technique in a quasi-relativistic framework, and we have benchmarked this
approach recently for CHBrClF, CHClFI, CHBrFI and CHAtFI \cite{bruck:2021}.
With this new approach, we are finally in the position to estimate
rotational frequency splittings in transition metal complexes, most
importantly the specific Re system analysed in high-resolution molecular
spectroscopy by Schnell and coworkers \cite{medcraft:2014}.

CpRe(CH$_{3}$)(CO)(NO) is quite interesting in several aspects: Firstly, the
presence of both naturally occurring rhenium isotopes ($^{187}$Re and
$^{185}$Re) and nitrogen ($^{14}$N) with nuclear spins $I \geq 1$
($I_\mathrm{Re} = 5/2$ and $I_\mathrm{N} = 1$) leads to significant hyperfine
structure in the spectrum. 
Secondly, the magnitudes of the nuclear quadrupole moments of the
rhenium and nitrogen nuclei
differ by more than two orders of magnitude. 
The magnitude of the hyperfine splitting reflects, besides the size of the
electric quadrupole moment of the specific isotope, also the anisotropy in the 
electric-field gradient surrounding the
quadrupolar nucleus, which can be used to describe the bonding
situation in the vicinity of the respective nuclei. 
Thus, NEQC is sensitive to the electronic density in the vicinity of
the nucleus, which is relevant for molecular parity violation, and thus 
having measurements of NEQC available allows to assess the quality of the 
theoretical description.

\begin{figure*}[ht!]
\begin{tabular}{llccc}
\toprule
\tiny
&&\multicolumn{3}{c}{Molecular structure}\\
&Method & ZORA-HF & ZORA-B3LYP & RECP-B3LYP\\
\midrule
\multirow{4}{*}{\raisebox{-2.5\height}{\rotatebox{90}{Electronic
density}}}&
 HF   &
\raisebox{-.5\height}{\includegraphics[width=0.2\textwidth,trim=1cm 1cm 1cm 1cm,clip]{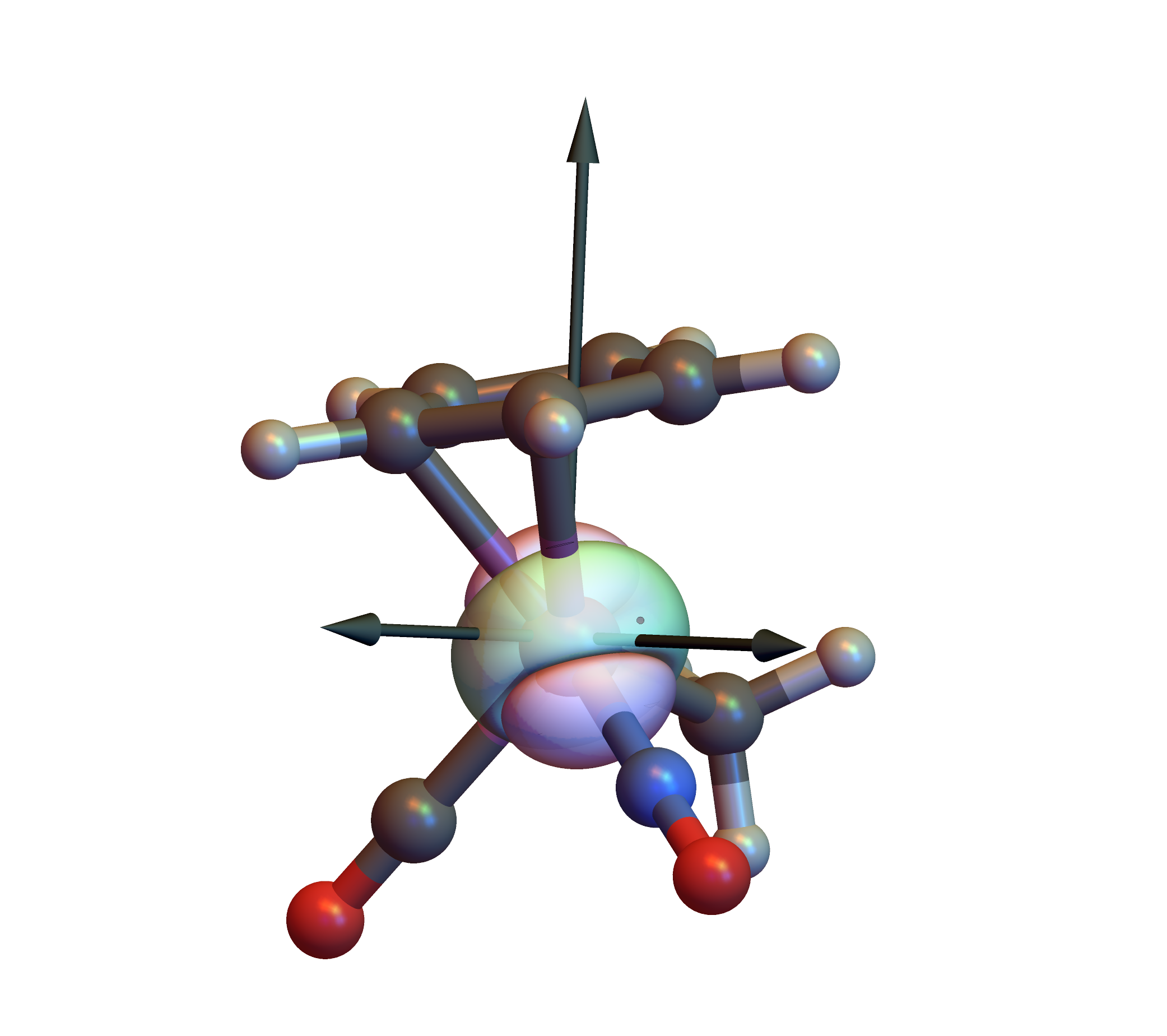}} &
\raisebox{-.5\height}{\includegraphics[width=0.2\textwidth,trim=1cm 1cm 1cm 1cm,clip]{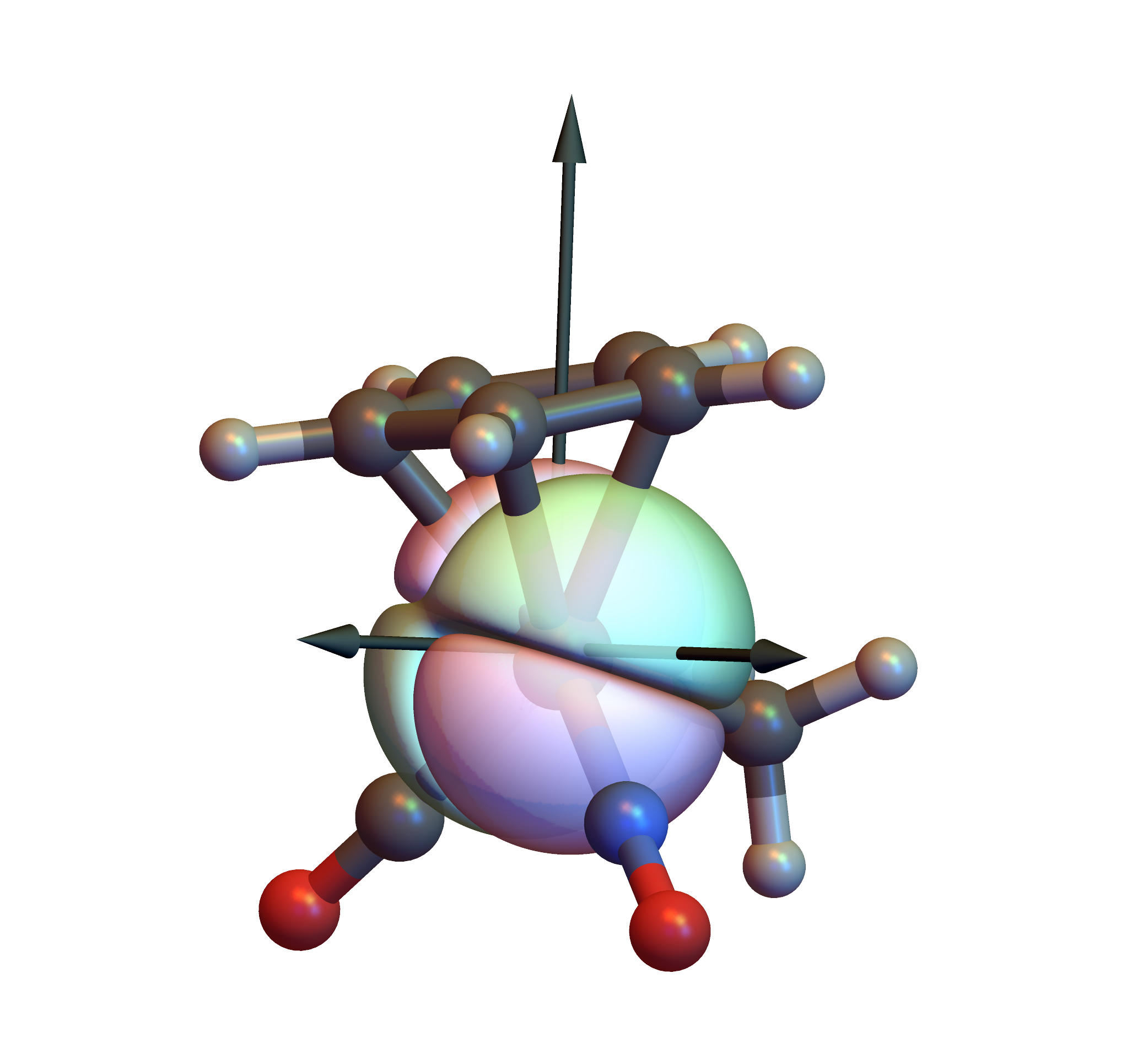}} &
\raisebox{-.5\height}{\includegraphics[width=0.2\textwidth,trim=1cm 1cm 1cm 1cm,clip]{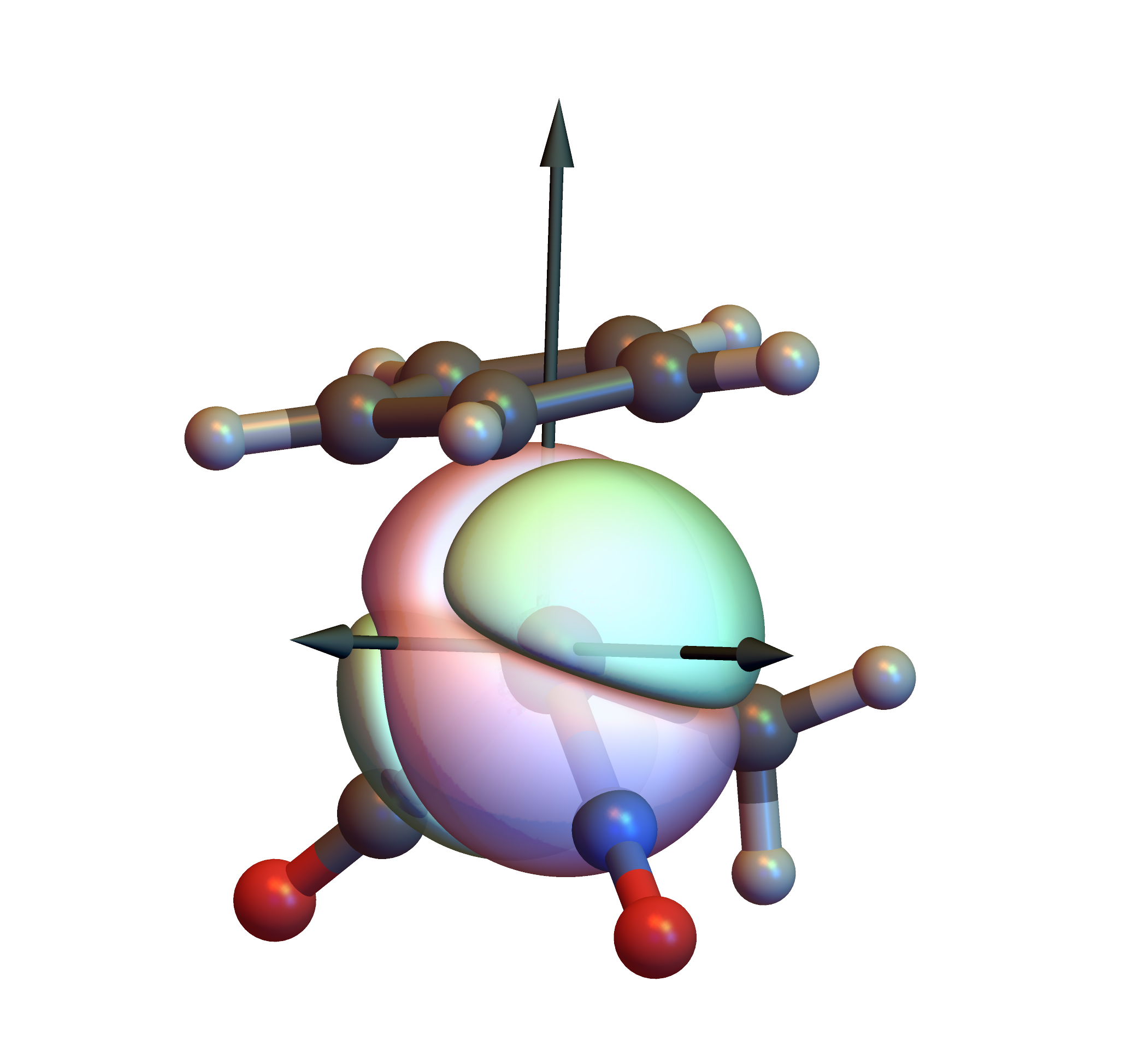}} \\
&B3LYP &  
\raisebox{-.5\height}{\includegraphics[width=0.2\textwidth,trim=1cm 1cm 1cm 1cm,clip]{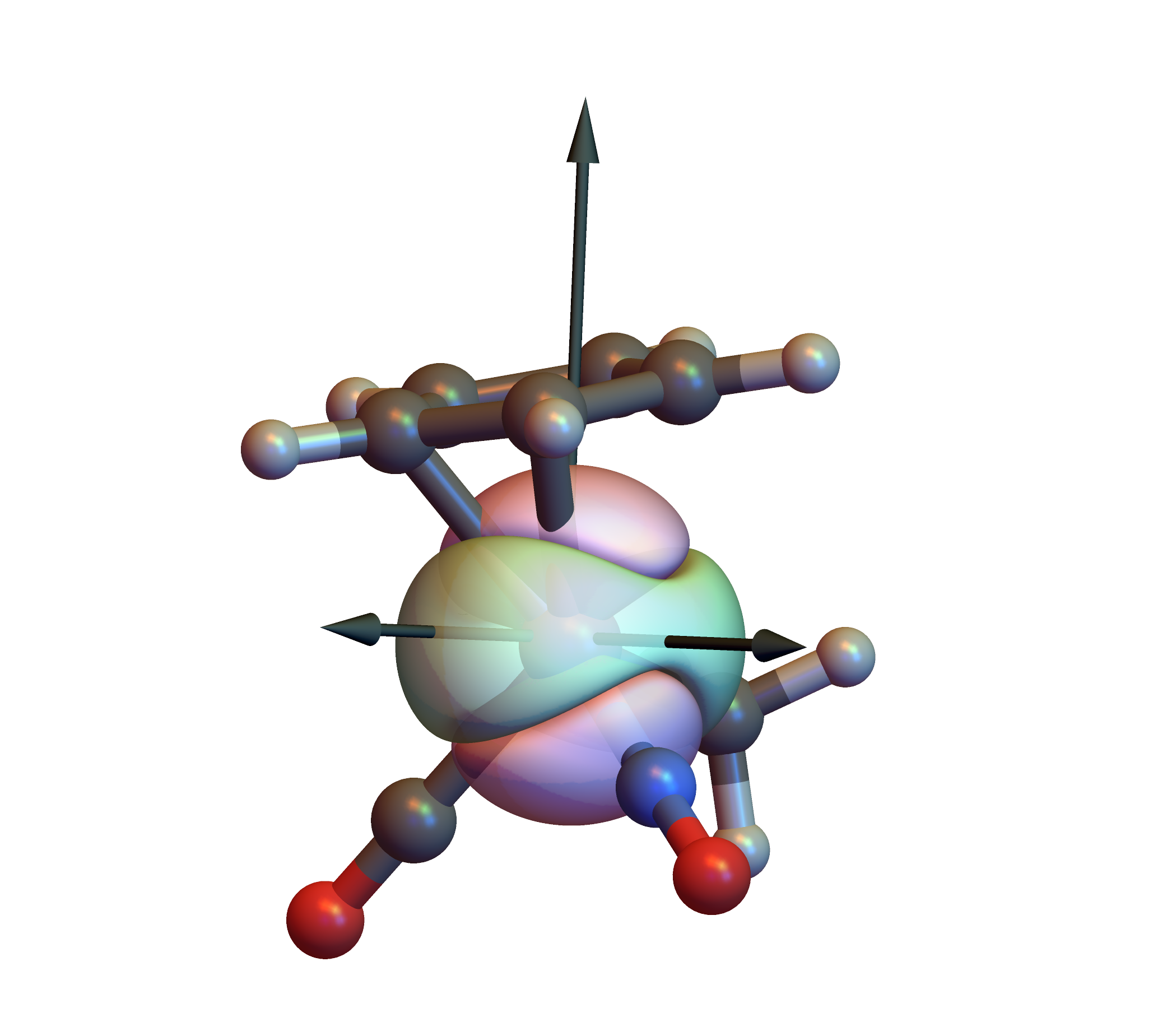}} &
\raisebox{-.5\height}{\includegraphics[width=0.2\textwidth,trim=1cm 1cm 1cm 1cm,clip]{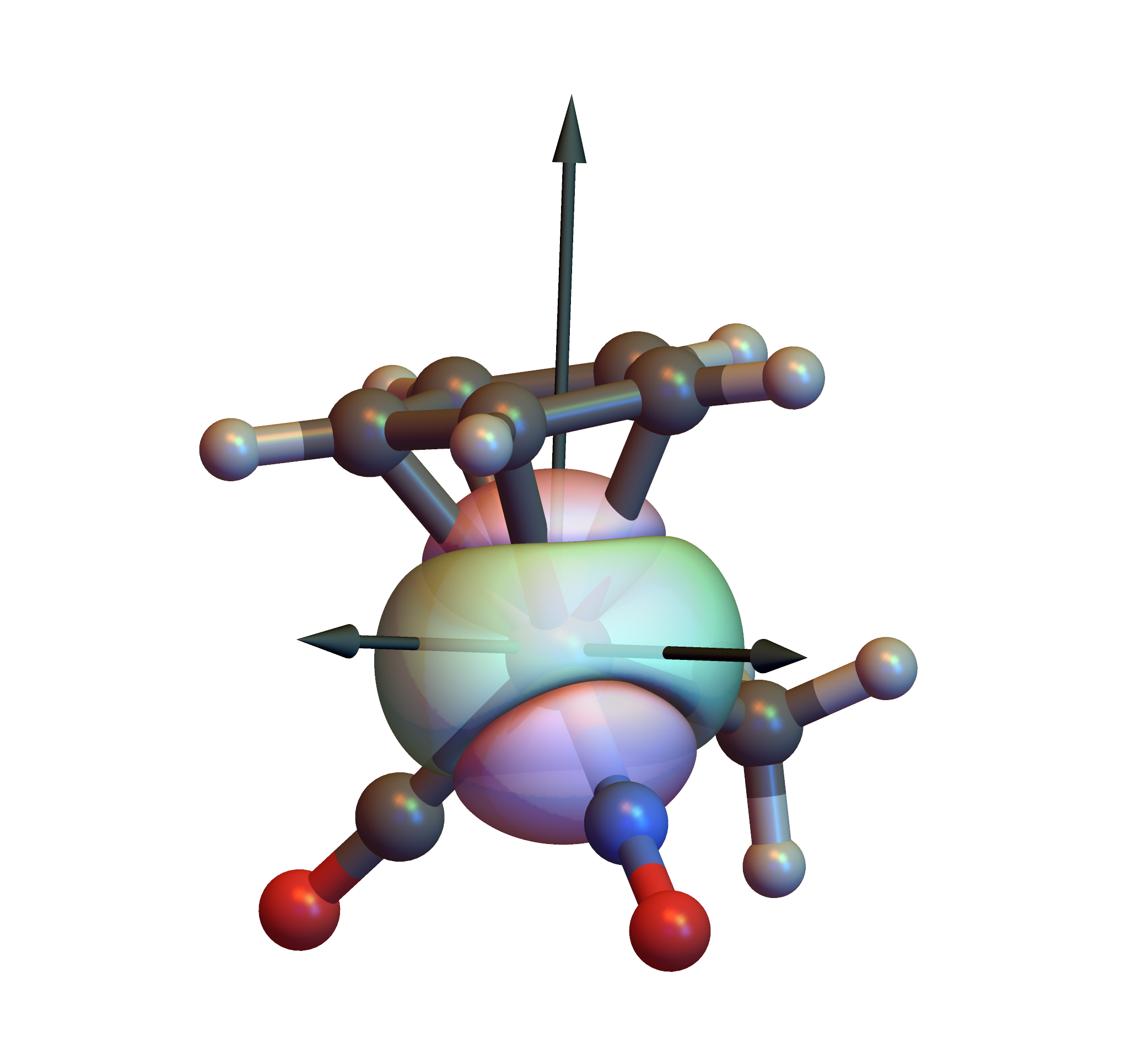}} &
\raisebox{-.5\height}{\includegraphics[width=0.2\textwidth,trim=1cm 1cm 1cm 1cm,clip]{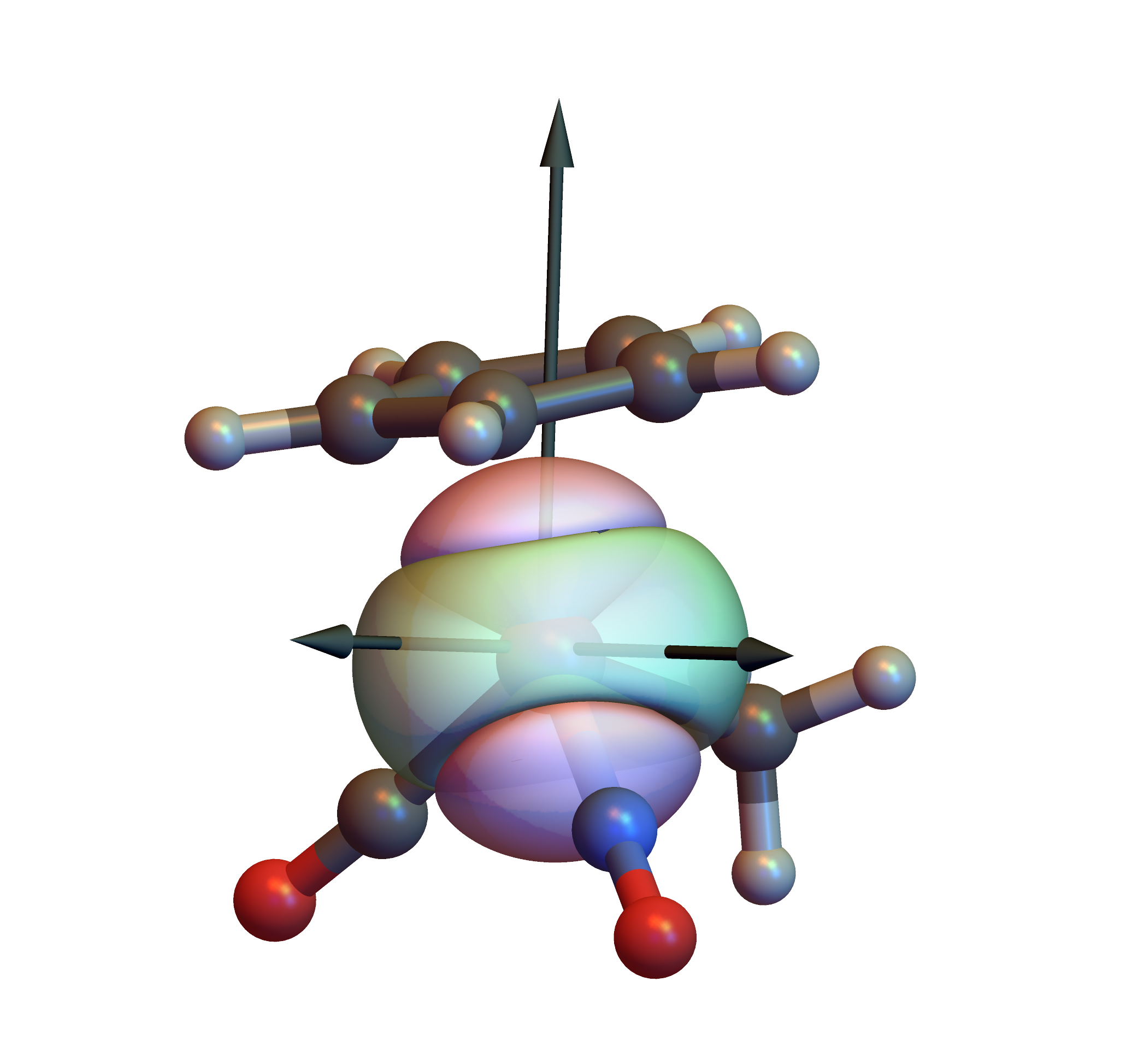}} \\
&LDA   &  
\raisebox{-.5\height}{\includegraphics[width=0.2\textwidth,trim=1cm 1cm 1cm 1cm,clip]{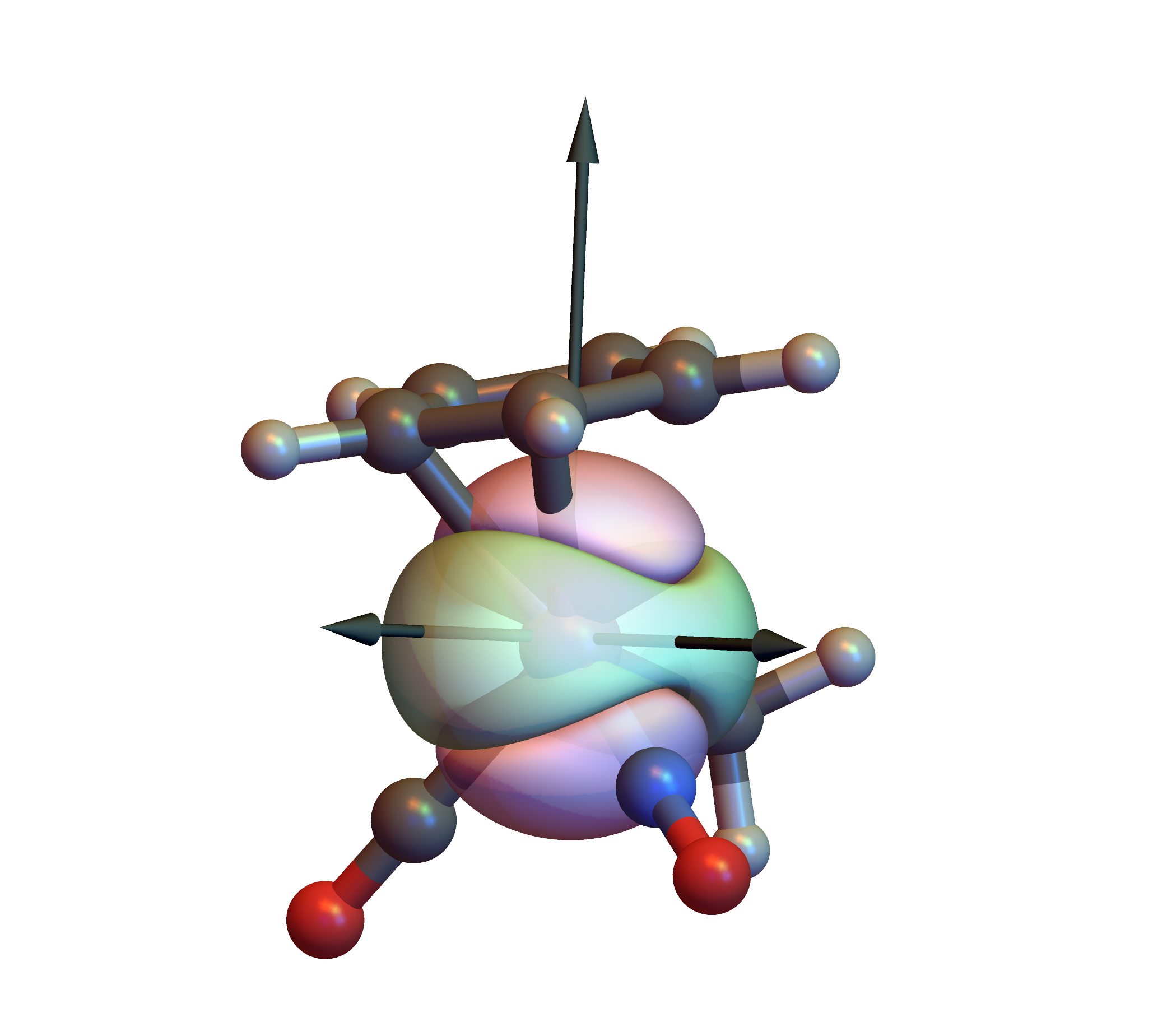}} &
\raisebox{-.5\height}{\includegraphics[width=0.2\textwidth,trim=1cm 1cm 1cm 1cm,clip]{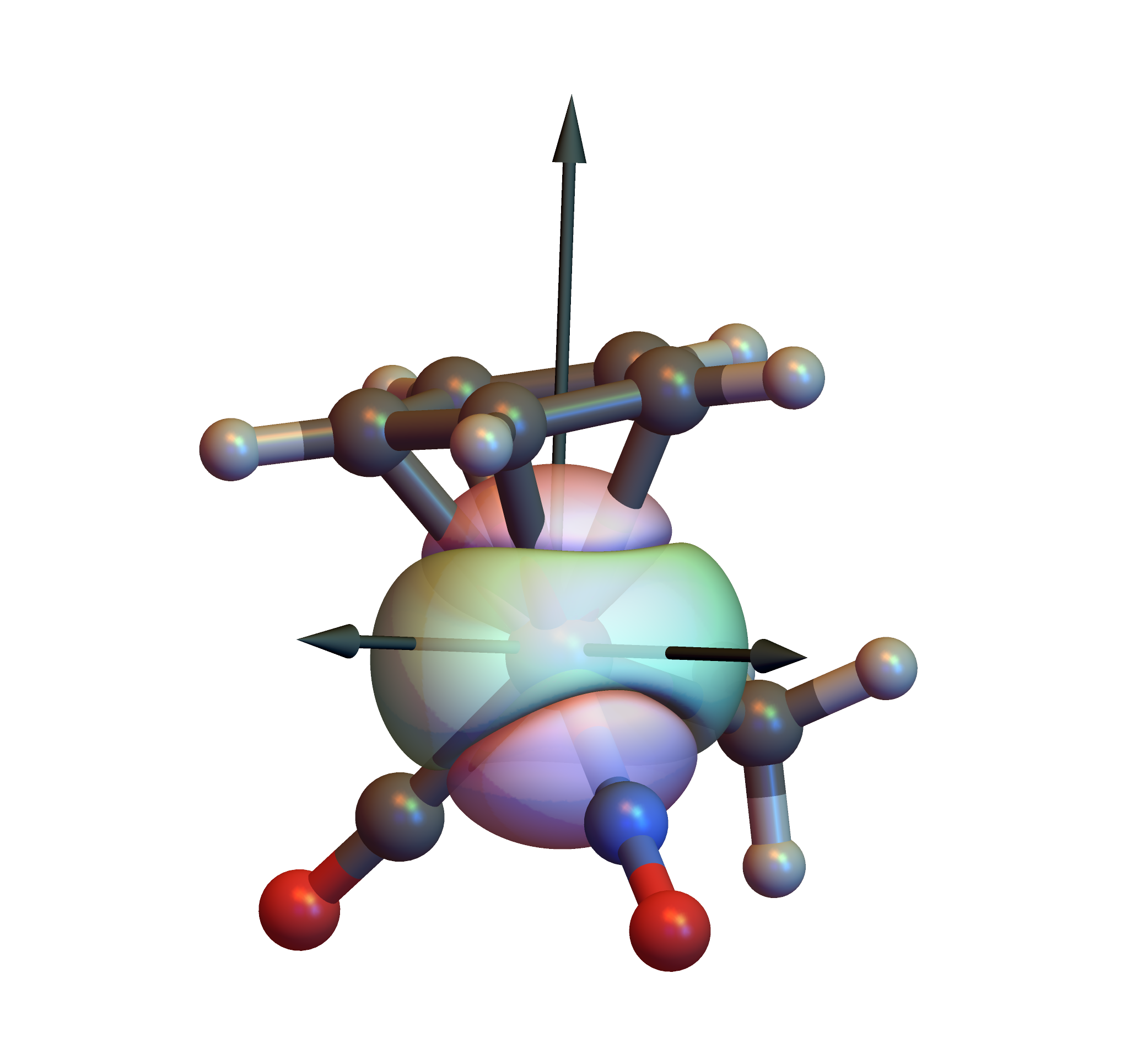}} &
\raisebox{-.5\height}{\includegraphics[width=0.2\textwidth,trim=1cm 1cm 1cm 1cm,clip]{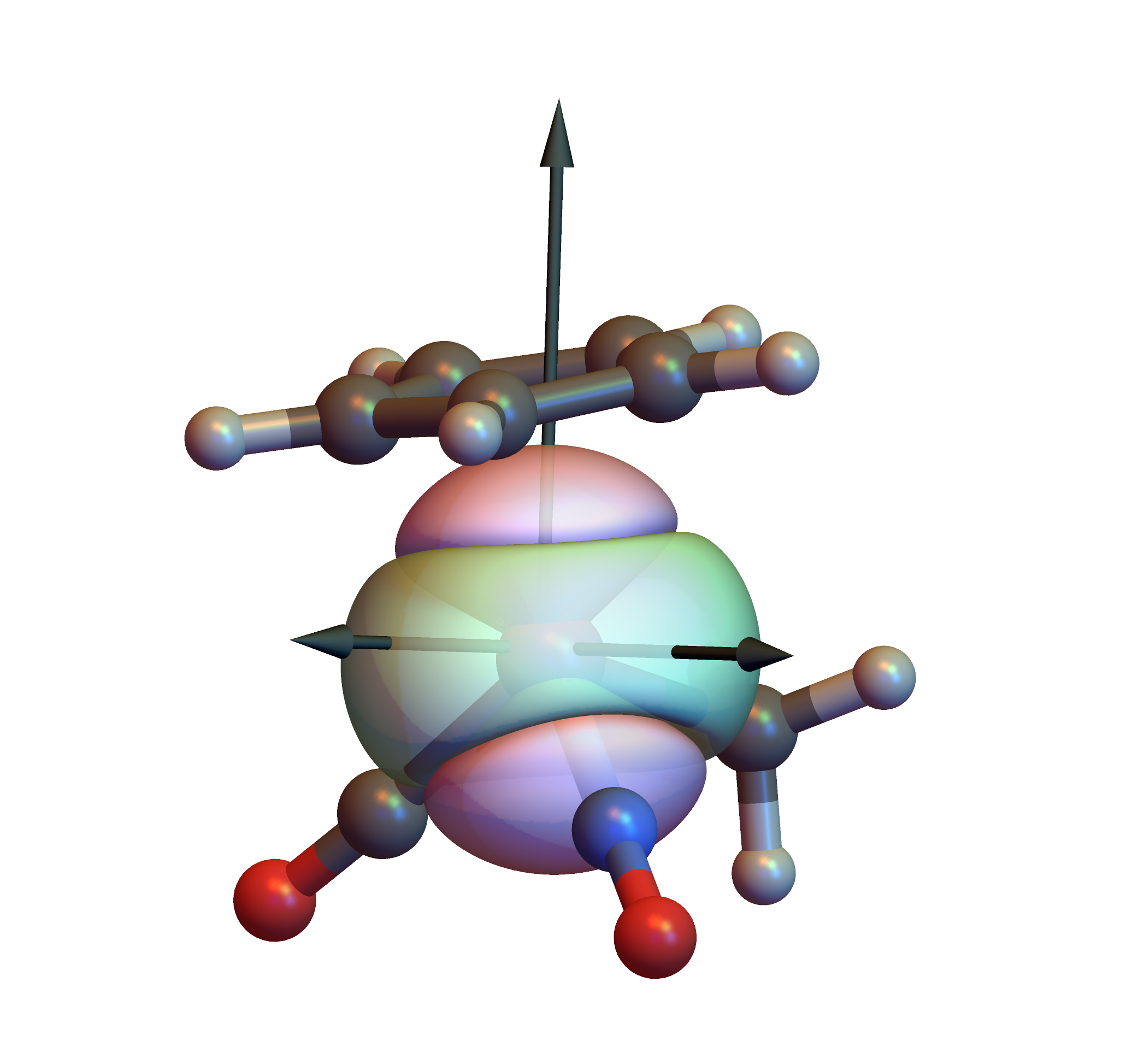}} \\
&Exp&  & \raisebox{-.5\height}{\includegraphics[width=0.2\textwidth,trim=1cm 1cm 1cm 1cm,clip]{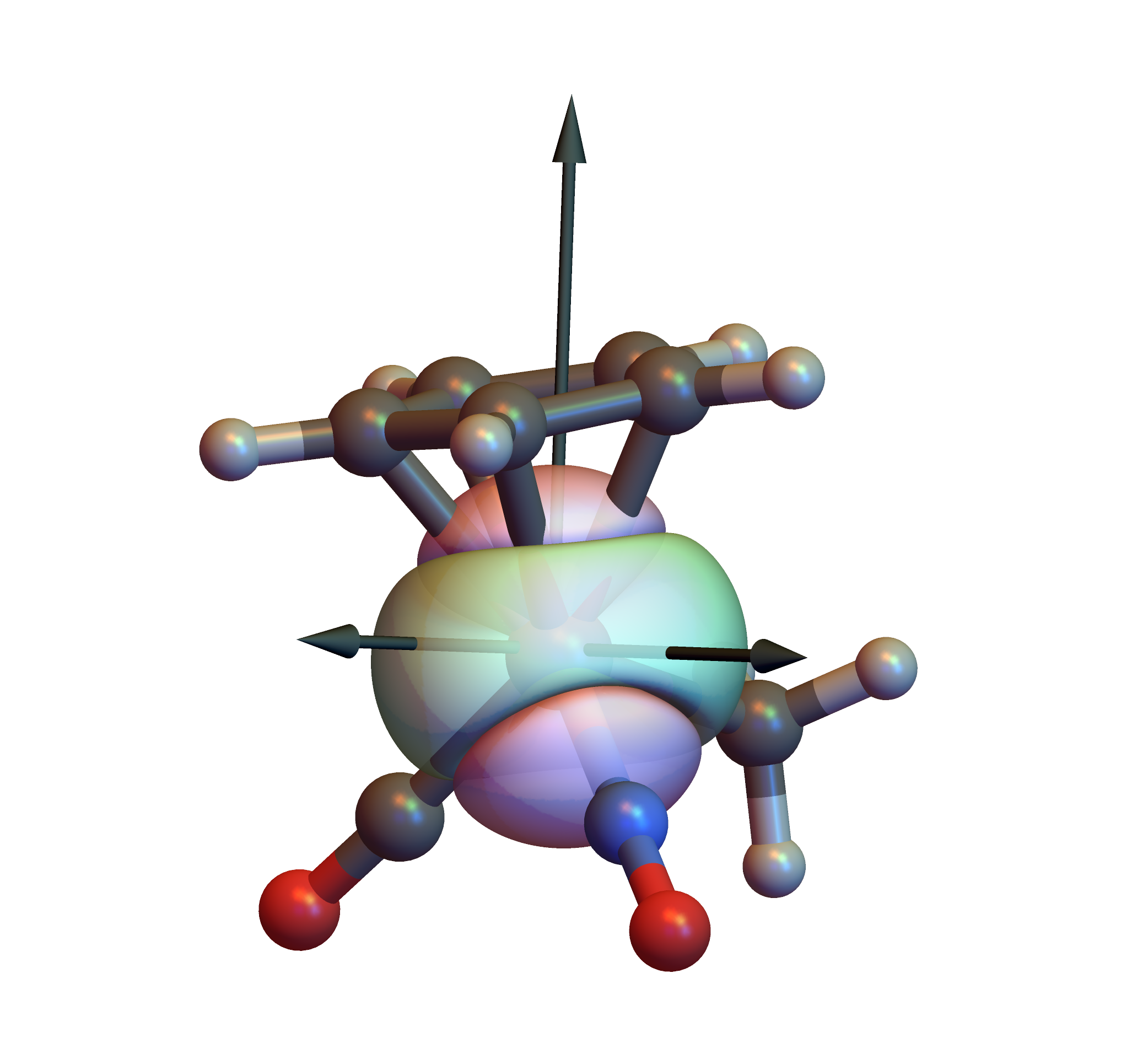}} \\
\bottomrule
\end{tabular}
\caption{Visualization of nuclear electric quadrupole coupling tensors
of $^{187}$Re in Cp$^{187}$Re(CH$_{3}$)(CO)(NO) with polar plots as
described in Ref.~\cite{autschbach:2010}. Values for the corresponding experimental
NEQC were taken from Ref.~\cite{medcraft:2014}. A nuclear electric
quadrupole coupling constant of $eQ=207.0\,\mathrm{fm}^2$ was used for
$^{187}$Re as suggested in Ref.~\cite{pyykko:2018}. Molecules are
aligned along the principal axes of rotation, which are shown as black
arrows. Lengths of these arrows are scaled by the value of the
corresponding rotational constant $A$, $B$, $C$.}
\label{fig: neqc}
\end{figure*}

We consider herein the {\it R}-enantiomer of CpRe(CH$_{3}$)(CO)(NO)
for our study ({\it cf.} Fig.~\ref{fig: neqc}).  Equilibrium
rotational constants calculated with relativistic effective core
potential (RECP) are listed in the Supporting Information and appear
at first glance to be in reasonable agreement with experimentally
observed values reported in Ref.~\cite{medcraft:2014}.  On closer
inspection, however, the RECP molecular structures obtained on the
B3LYP level agree extremely well with regard to the measured $A$
constant, but have smaller $B$ and $C$ constants because the
cyclopentadienyl ring is found at too large a distance from the
rhenium atom on this level. This structural feature impacts
significantly on the computed NEQC constants in
Cp$^{187}$Re(CH$_{3}$)(CO)(NO), for which we used a quasi-relativistic
zeroth-order regular approximation (ZORA) approach with different
density functionals (ZORA-DFT) or within a Hartree--Fock framework
(ZORA-HF) as described in Refs.~\cite{gaul:2020,gaul:2020b}. We
compare to the NEQC tensor experimentally determined in Ref.
\cite{medcraft:2014} and study the influence of different density
functionals on the energy optimization of the molecular structure as
well as on molecular electronic density. A visualization of selected
results for the NEQC tensor of $^{187}$Re in a polar tensor plot as suggested in Ref.
\cite{autschbach:2010} is shown in Fig.~\ref{fig: neqc}, whereas a complete
list of results for the full NEQC tensors for $^{187}$Re and $^{14}$N at different levels of theory
is provided in the Supplement. One can clearly see from Fig.~\ref{fig: neqc} 
that HF gives rise to qualitatively wrong NEQC tensors both for the electronic 
density as well as for the molecular structure. Best
results are received with hybrid functionals such as B3LYP and
molecular structures optimized at the ZORA-DFT level. The difference
between different functionals for computation of the electron density
is comparatively small, in particular we observe that LDA agrees with
B3LYP within $10\%$ for the components of the NEQC tensor. This
finding suggests that also for the prediction of PV effects in this
chiral Re complex, results at the ZORA-DFT level are to be favoured
over those from a HF treatment as the latter yields a qualitatively
wrong description of the electron density close to the rhenium
nucleus. To be specific for the parity conserving potentials, we use
in the following the two-component ZORA-B3LYP equilibrium structure
reported in Table S1 together with the corresponding harmonic
vibrational force fields of Table S9 and S10. This
structure provides on average the lowest deviations of the NEQC tensor
from the experimental data, and various density functionals result in
similar deviations, even when using the LDA functional. 

After these benchmarks, we calculated with our two-component ZORA approach to electroweak quantum chemistry
\cite{berger:2005,berger:2005a,nahrwold:09,gaul:2020,bruck:2021} 
the molecular PV energy $E_{\mathrm{PV}}^{R}$ of the {\it
R}-CpRe(CH$_{3}$)(CO)(NO) enantiomer displayed on the right in Fig 1.
For the ZORA-B3LYP as well as the RECP-B3LYP equilibrium structure,
the values of $E_{\mathrm{PV}}^{R}$ for both of the isotopologues are
of the order of 10$^{-9}$ cm$^{-1}~hc$ ({\it cf.} Fig. 3). The
difference of the PV potentials [$\Delta E_{\mathrm{PV}}^{R,S} = 2
\times E_{\mathrm{PV}}^{R}$] between the {\it R}- and {\it
S}-enantiomers is about 3 $\times$ 10$^{-9}$ cm$^{-1}~hc$ (ca.
90~Hz~$h$ or 40~nJ~mol$^{-1}$).  These effect sizes are comparable to
previous predictions for other rhenium complexes
\cite{schwerdtfeger:2003,schwerdtfeger:2004b,darquie:2010}. The values
of $E_{\mathrm{PV}}^{R}$ are seen to depend on the choice of level of
theory where both LDA and BLYP values are close to each other,
deviating by about 1~\% from each other, whereas B3LYP produces
similar values and is numerically $\approx$ 15$\%$ larger than LDA and
BLYP ones (see Supplementary Material). However, the HF
$E_{\mathrm{PV}}^{R}$ value depends strongly on the molecular
structure, being $\approx$ 15$\%$ larger than the LDA value for the
RECP-B3LYP structure but $\approx$ 15$\%$ smaller than the LDA value
for the ZORA-B3LYP structure

\begin{figure*}
\includegraphics[width=\textwidth]{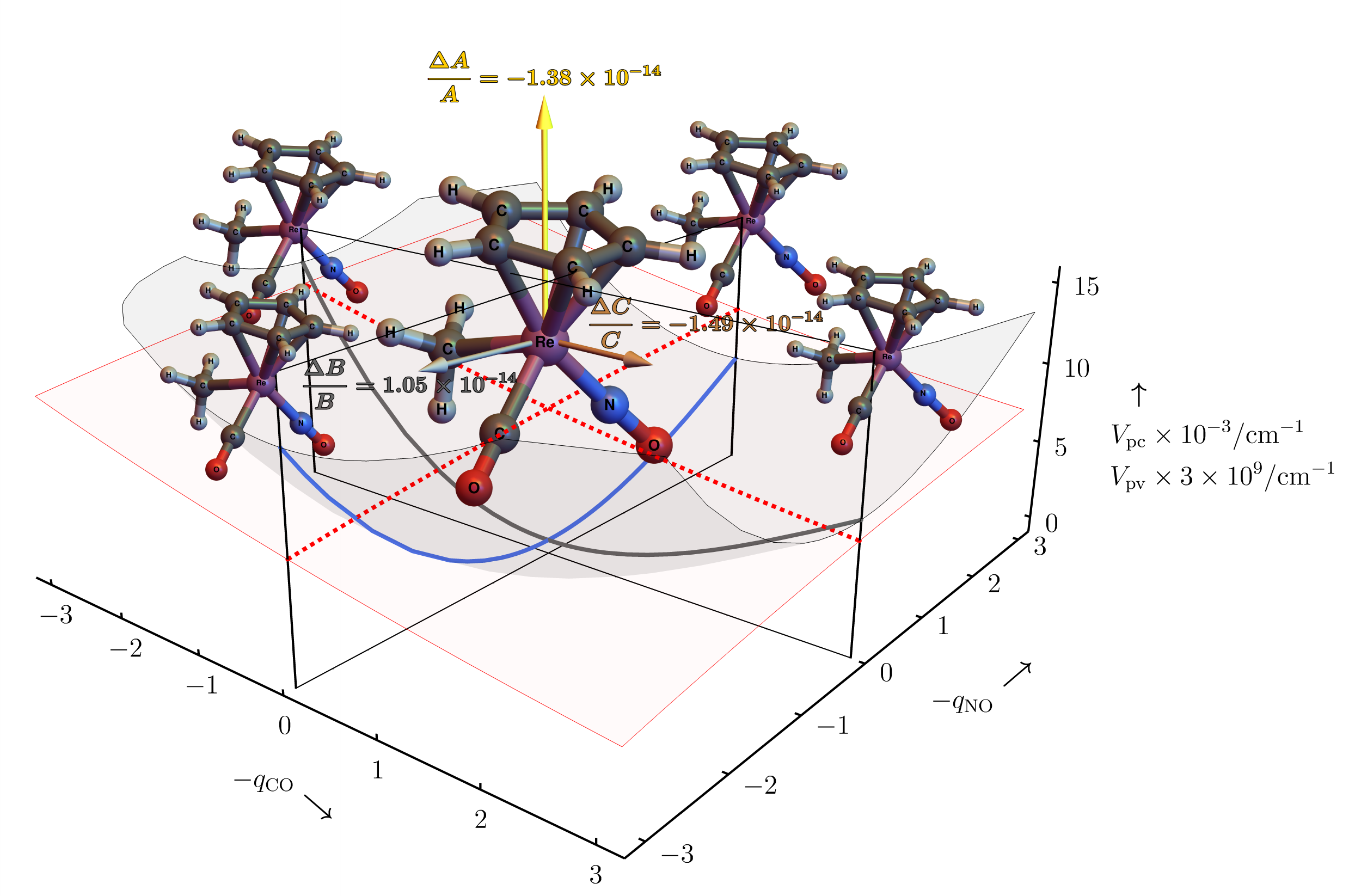}
\caption{Visualization of parity violating shifts of the rotational
constants of $R$-Cp$^{187}$Re(CH$_{3}$)(CO)(NO) as computed on the
level of ZORA-LDA with the molecular structure optimized at the level
of ZORA-B3LYP/x2c-TZVPPall-2c. The principal axes of rotation are
shown in gold, silver and bronze with their length scaled by the value
of the corresponding rotational constant $A$, $B$, $C$ alongside their
parity violating relative shifts $\Delta A/A=-1.38\times10^{-14}$,
$\Delta B/B=1.05\times10^{-14}$ and $\Delta C/C=-1.49\times10^{-14}$.
The parity violating shifts of the rotational constants are
proportional to the gradient of the parity violating potential at the
equilibrium structure. The parity conserving potential $V_\mathrm{pc}$
is symbolically shown for the NO and CO stretching modes
($q_\mathrm{NO}$ and $q_\mathrm{CO}$) in the color of the involved
atom (N blue and C gray). The parity violating potential is shown
along these two modes as dotted red line. Surfaces that connect the
cuts along $q_\mathrm{NO}$ and $q_\mathrm{CO}$ are shown to guide the
eye but were not explicitly computed.}
\label{fig:Results}
\end{figure*}

The experimental infrared spectrum of the complex shows two very strong
signals at 1664 and 1932 cm$^{-1}$ \cite{medcraft:2014}, which correspond to
the NO and CO stretching fundamentals, respectively. The gas-phase
harmonic force field computed at the two-component ZORA-B3LYP
level for {\it R}-CpRe(CH$_{3}$)(CO)(NO) for both of
the $^{187}$Re (natural abundance 62.6 $\%$) and $^{185}$Re (natural
abundance 34.7 $\%$) isotopologues give unscaled transition wavenumbers at
1791 cm$^{-1}$ for the NO stretch and 2032 cm$^{-1}$ for the CO stretch
vibration. 
The variation of parity conserving (PC) and PV potentials as a function of the dimensionless 
reduced normal coordinate corresponding to CO and NO stretching modes of 
Cp$^{187}$Re(CH$_{3}$)(CO)(NO) are displayed in Fig. 3. The former potential is
obtained at the two-component ZORA-B3LYP level, the latter is evaluated
within a two-component ZORA-LDA approach.
Cuts through PV potentials along the dimensionless reduced normal
coordinates for other functionals like
B3LYP and BLYP (see Supporting Information) follow similar patterns as that of LDA, whereas the 
plot due to HF is again different and deviates from the DFT methods, 
indicating, as already observed in Fig.~\ref{fig: neqc}, that 
HF results are unreliable for describing electronic properties of compounds containing
transition metal centers.  
The LDA-slope of the PV potential in this two-dimensional vibrational manifold 
induces a tiny parity-violating change of the equilibrium structure 
leading to a shortened NO bond and an elongated CO bond of the R-enantiomer and
vice versa for its mirror image.

The presence of non-zero PV energy gradients $\vec \nabla E_\mathrm{PV}$ at 
the minimum of the PC potential ($\it{cf.}$ Tables
S3-S4) leads, thus, to a change 
of the equilibrium structure of a chiral molecule.
Inspired by the existing experimental broadband rotational spectrum of the CpRe(CH$_{3}$)(CO)(NO) 
complex \cite{medcraft:2014}, the analytically derived Cartesian PV energy gradients $\vec \nabla E_\mathrm{PV}$ 
are used to estimate the PV induced shifts of the 
equilibrium rotational constants.
For the ${}^{187}$Re isotopologue, the values of $\frac{\Delta X_\mathrm R}{X_\mathrm R}$
as obtained on the LDA level are displayed in Fig. 3, other data can be
located in the Supplementary Material. The relative shifts are
favourably large, of the order of 10$^{-14}$  for all 
rotational constants. A similar magnitude but opposite signs were obtained
for these on the HF level, but these latter values are to be discarded
by virtue of the erroneous HF results for NEQCs in this compound. 

Our explicit predictions for PV in rotational transitions of a 5d
transition metal compound thus support Letokhov's rule of thumb for an
overall consistency of relative PV effect sizes in rotational and
vibrational transitions. In vibrational transitions, $\Delta \nu/\nu
\approx 10^{-14}$ are projected to be resolvable with present day
techniques for tailored experimental setups \cite{cournol:2019}. In
rotational spectroscopy, however, the relative experimental resolution is 
presently lower for standard experiments, but here one can strongly benefit 
from a linear scaling of the splittings with rotational quantum number $J$, 
so that orders of magnitude can in principle be gained in high $J$ 
transitions \cite{quack:2000}. The resolution of a typical coaxial Fourier 
transform microwave spectroscopy experiment \cite{grabow:2005} employing 
a microwave resonator combined with supersonic expansion of the
molecular sample is currently limited to 3--10~kHz ($\approx 10^{-7}$ 
on the relative scale) depending on the carrier
gas and thus on the velocity of the molecular beam due to mainly two
reasons: i) the limited flight time of the molecules in the resonator and
ii) the transverse motion of the molecules during the supersonic expansion
accompanied by a significant loss of molecules from the area of high
microwave field strength in the resonator \cite{schnell:2011}. The flight 
time can be increased significantly for example by elongating the resonator 
distance $d$ and by using sources of translationally cold, decelerated 
molecules. For 20~m~s$^{-1}$ slow molecules \cite{meerakker:2008} and 
$d = 1~\mathrm{m}$, the molecules can be monitored up to 50~ms compared to
1~ms for non-decelerated beams using neon as a carrier gas
(1000~m~s$^{-1}$).~\cite{schnell:2011} The resolution will be limited to
20--100~Hz Doppler broadening, but when measuring frequency differences
between enantiomers in an alternating way, the resolving power of the
experiment is not limited by the linewidth, but rather by the precision
with which the frequency of the line centers can be determined. This is
usually about one to two orders of magnitude better than the 
linewidths~\cite{quack:2008}, so that the Hz range should be
reachable. Furthermore, enantiomer-selective population transfer experiments 
based on advanced microwave pulse schemes can selectively populate or 
depopulate certain rotational states of interests by just changing the 
phase of one microwave pulse in the scheme, which can be highly 
advantageous for such experiments.~\cite{eibenberger:2017,singh:2023} Alternative 
approaches to perform microwave spectroscopy with perspectively even higher resolution
are Ramsey type arrangements using the method of separated oscillatory 
fields.~\cite{nijs:2014}

We note in passing that we also attempted to predict parity violating
wavenumber splittings in the CO and NO stretching fundamental using our
recently described perturbative approach that uses finite differences of
analytic PV potential gradients together with cubic parity conserving
potentials to take multi-mode effects into account \cite{bruck:2021}.
Although again relative PV splittings on the order of $10^{-14}$ are
obtained, which seem to be very reasonable estimates, the values change
sign when multi-mode effects are accounted for. Due to near resonance
conditions that also lead to significant IR intensity redistribution in the
NO and CO stretching fundamental, a more detailed analysis of the coupling
situation seems mandatory. Nevertheless, we report corresponding data as
obtained for the slightly inferior RECP-B3LYP optimised structures and
their (an)harmonic force fields in the Supplement for future reference.

In summary, we were able to predict with our recently established
quasi-relativistic methodology \cite{bruck:2021} the relative
shifts in rotational constants that are induced by electroweak parity
violation in the equilibrium structure of 
the heavy-elemental chiral complex CpRe(CH$_{3}$)(CO)(NO) containing either $^{187}$Re or $^{185}$Re isotopes, for which 
a high-resolution gas-phase rotational spectrum has already been reported \cite{medcraft:2014}. 
Analytically derived $\vec \nabla E_\mathrm{PV}$ have been utilised 
to predict large relative shifts of rotational constants of about 10$^{-14}$.
Due to pronounced barrier heights for internal 
rotation of the Cp and methyl groups, the chiral dynamics of
CpRe(CH$_{3}$)(CO)(NO) in the rovibronic ground state is
expected to be dominated, similarly to \ce{S2Cl2} \cite{berger:2005,sahu:2021}, by parity violating effects rather than tunnelling phenomena.
The comparatively large PV effect in this compound is excellent news for
experimental attempts to measure for the first time molecular parity violation
in precision rotational spectroscopy of chiral molecules.\\

%%%%%%%%%%%%%%%%%%%%%%%%%%%%%%%%%%%%%%%%%%%%%%%%%%%%%%%%%%%%%%%%%%%%%%%%%%%%%

{\large \bf \noindent \it
\fontfamily{phv}\selectfont Computational Details} \\

Molecular structures were optimized at different levels of theory
using the turbomole program package version 7.5\cite{turbomole7.5} for
non-relativistic calculations with a relativistic effective core
potential (RECP) and a modified version \cite{wullen:2010} of the
program package turbomole \cite{ahlrichs:1989} for two-component ZORA
calculations. Molecular structures were optimized at the level of DFT
employing the PBE exchange-correlation functional \cite{perdew:1996},
its hybrid version PBE0~\cite{adamo:1999} or the B3LYP
exchange-correlation
functional~\cite{becke:1988,lee:1988,vosko:1980,becke:1993} and on the
level of HF. In RECP calculations, an Ahlrichs basis set of triple zeta
quality with additional polarisation functions def2-TZVPP
\cite{schafer:1994,weigend:1998} was
used for all atoms or a correlation consistent Dunning basis set of
quintuple zeta quality with additional polarisation functions cc-pV5Z
for H, C, N and O together with a correlation consistent quadruple zeta
basis set with core valence polarisation cc-pwCVQZ on Re. In
both cases, an energy-consistent RECP of the Stuttgart group was
employed for Re \cite{figgen:2009}. In two-component ZORA calculations
a basis of triple zeta quality optimized for two-component
all-electron calculations x2c-TZVPPall-2c was used.\cite{pollak:2017}
Molecular structures were optimized until the norm of the cartesian
gradient was below $10^{-4}\,E_\mathrm{h}\,a_0^{-1}$. 

For NEQC calculations, electronic densities were optimized for all
molecular structures on the two-component ZORA level with the
x2c-TZVPPall-2c basis on H, C, N and O and an even-tempered basis set
with 25 s, 25 p, 14 d, 11 f and 3 g functions on Re.  The parameters
of the even-tempered series are composed as $\alpha_i= \gamma
\beta_N^{N-i}, \, i=1,\dots,N$, with $N=26$, $\gamma=0.02$ and
$\alpha_1=500000000$, where exponents $\alpha_{1-25}$ were used for s
functions, exponents $\alpha_{2-26}$ were used for p functions,
exponents $\alpha_{12-25}$ were used for d functions, exponents
$\alpha_{15-25}$ were used for f functions and exponents
$\alpha_{21-23}$ were used for g functions. All calculations were
converged until energy changes between two successive iterations were
smaller than $10^{-9}\,E_\mathrm{h}$. Electronic densities were
computed at the level of HF, PBE, PBE0 and B3LYP as well as with the
DFT exchange-correlation functionals LDA\cite{slater:1951,vosko:1980},
BLYP\cite{becke:1988,lee:1988,vosko:1980} and BHLYP.

For calculations of PV energies and PV gradients, electronic densities
were optimized for the molecular structures on the level of
RECP-B3LYP/cc-pV5Z,cc-pwCVQZ-PP (Ref.~\cite{medcraft:2014}) and
ZORA-B3LYP/x2c-TZVPPall-2c, on the two-component ZORA level with the
aug-cc-pVDZ basis set for hydrogen and the even-tempered basis set
described above utilising  for s and p functions, the exponents
$\alpha_{1-25}$ and $\alpha_{2-26}$ and in the case of d functions
$\alpha_{20-24}$ for H, C, N and O atoms, and $\alpha_{12-25}$ for the
Re atom.  Additionally, for Re f functions with exponents
$\alpha_{15-22}$ were used. PV energies and gradients were computed at
the level of HF, B3LYP, BLYP and LDA. All calculations were converged
until energy changes were smaller than $10^{-9}$~$E_{\textrm{h}}$
whereas the relative change of spin-orbit energy between
two-successive iterations were dropped below $10^{-13}$ . The
threshold for neglection of gradients of two-electron integrals was
set to $10^{-15}$~$E_{\textrm{h}}\,a_{0}^{-1}$. 
 
In all two-component calculations of electronic densities and PV
gradients, a normalized spherical Gaussian nuclear charge density
distribution $\rho_A \left( \vec{r} \right) = \frac{\zeta_A^{3/2}}{\pi
^{3/2}} \text{e}^{-\zeta_A \left| \vec{r} - \vec{r}_A \right| ^2}$ was
used with $\zeta_A = \frac{3}{2 r^2 _{\text{nuc},A}}$ and the
root-mean-square radius $r_{\text{nuc},A}$ was chosen as suggested by
Visscher and Dyall \cite{visscher:1997}. Unless explicit isotopes are
given, the nuclear mass number was determined as nearest integer to the
relative atomic mass. 

NEQC tensors and PV energies of two-component ZORA densities were computed
using our toolbox approach \cite{gaul:2020}. Within this, NEQC tensors
were computed as described in Ref. \cite{gaul:2020b}.

The analytically derived cartesian PV energy gradient 
$\vec \nabla E_\mathrm{PV}$ at the equilibrium structure 
of R-CpRe(CH$_{3}$)(CO)(NO) was computed as described in
Ref.~\cite{bruck:2021} and utilized for the estimation
of the relative shifts of the rotational constants \cite{quack:2000}, 
which are in principle associated with the rotational transitions as
described in Refs.~\cite{quack:2000a,bruck:2021}. 

\providecommand*{\mcitethebibliography}{\thebibliography}
\csname @ifundefined\endcsname{endmcitethebibliography}
{\let\endmcitethebibliography\endthebibliography}{}


\begin{mcitethebibliography}{67}
\providecommand*{\natexlab}[1]{#1}
\providecommand*{\mciteSetBstSublistMode}[1]{}
\providecommand*{\mciteSetBstMaxWidthForm}[2]{}
\providecommand*{\mciteBstWouldAddEndPuncttrue}
  {\def\EndOfBibitem{\unskip.}}
\providecommand*{\mciteBstWouldAddEndPunctfalse}
  {\let\EndOfBibitem\relax}
\providecommand*{\mciteSetBstMidEndSepPunct}[3]{}
\providecommand*{\mciteSetBstSublistLabelBeginEnd}[3]{}
\providecommand*{\EndOfBibitem}{}
\mciteSetBstSublistMode{f}
\mciteSetBstMaxWidthForm{subitem}
{\alph{mcitesubitemcount})}
\mciteSetBstSublistLabelBeginEnd{\mcitemaxwidthsubitemform\space}
{\relax}{\relax}

\bibitem[Berger and Stohner(2019)]{berger:2019}
R.~Berger, J.~Stohner, \emph{Wiley Interdiscip. Rev.-Comput. Mol. Sci.}
  \textbf{2019}, \emph{9}, e1396\relax
\mciteBstWouldAddEndPuncttrue
\mciteSetBstMidEndSepPunct{\mcitedefaultmidpunct}
{\mcitedefaultendpunct}{\mcitedefaultseppunct}\relax
\EndOfBibitem
\bibitem[Yamagata(1966)]{yamagata:1966}
Y.~Yamagata, \emph{J. Theor. Biol.} \textbf{1966}, \emph{11}, 495--498\relax
\mciteBstWouldAddEndPuncttrue
\mciteSetBstMidEndSepPunct{\mcitedefaultmidpunct}
{\mcitedefaultendpunct}{\mcitedefaultseppunct}\relax
\EndOfBibitem
\bibitem[Letokhov(1975)]{letokhov:1975}
V.~S. Letokhov, \emph{Phys. Lett. A} \textbf{1975}, \emph{53}, 275--276\relax
\mciteBstWouldAddEndPuncttrue
\mciteSetBstMidEndSepPunct{\mcitedefaultmidpunct}
{\mcitedefaultendpunct}{\mcitedefaultseppunct}\relax
\EndOfBibitem
\bibitem[Zel'dovich \emph{et~al.}(1977)Zel'dovich, Saakyan, and
  Sobel'man]{zeldovich:1977}
B.~Y. Zel'dovich, D.~B. Saakyan, I.~I. Sobel'man, \emph{JETP Lett.}
  \textbf{1977}, \emph{25}, 94--97\relax
\mciteBstWouldAddEndPuncttrue
\mciteSetBstMidEndSepPunct{\mcitedefaultmidpunct}
{\mcitedefaultendpunct}{\mcitedefaultseppunct}\relax
\EndOfBibitem
\bibitem[Quack(1989)]{quack:1989}
M.~Quack, \emph{Angew. Chem. Int. Ed.} \textbf{1989}, \emph{28}, 571--586\relax
\mciteBstWouldAddEndPuncttrue
\mciteSetBstMidEndSepPunct{\mcitedefaultmidpunct}
{\mcitedefaultendpunct}{\mcitedefaultseppunct}\relax
\EndOfBibitem
\bibitem[Quack(2002)]{quack:2002}
M.~Quack, \emph{Angew. Chem. Int. Ed.} \textbf{2002}, \emph{41},
  4618--4630\relax
\mciteBstWouldAddEndPuncttrue
\mciteSetBstMidEndSepPunct{\mcitedefaultmidpunct}
{\mcitedefaultendpunct}{\mcitedefaultseppunct}\relax
\EndOfBibitem
\bibitem[Gaul \emph{et~al.}(2020)Gaul, Kozlov, Isaev, and Berger]{gaul:2020c}
K.~Gaul, M.~G. Kozlov, T.~A. Isaev, R.~Berger, \emph{Phys. Rev. Lett.}
  \textbf{2020}, \emph{125}, 123004\relax
\mciteBstWouldAddEndPuncttrue
\mciteSetBstMidEndSepPunct{\mcitedefaultmidpunct}
{\mcitedefaultendpunct}{\mcitedefaultseppunct}\relax
\EndOfBibitem
\bibitem[Gaul \emph{et~al.}(2020)Gaul, Kozlov, Isaev, and Berger]{gaul:2020d}
K.~Gaul, M.~G. Kozlov, T.~A. Isaev, R.~Berger, \emph{Phys. Rev. A}
  \textbf{2020}, \emph{102}, 032816\relax
\mciteBstWouldAddEndPuncttrue
\mciteSetBstMidEndSepPunct{\mcitedefaultmidpunct}
{\mcitedefaultendpunct}{\mcitedefaultseppunct}\relax
\EndOfBibitem
\bibitem[Kompanets \emph{et~al.}(1976)Kompanets, Kukudzhanov, Letokhov, and
  Gervits]{kompanets:1976}
O.~N. Kompanets, A.~R. Kukudzhanov, V.~S. Letokhov, L.~L. Gervits, \emph{Opt.
  Commun.} \textbf{1976}, \emph{19}, 414--416\relax
\mciteBstWouldAddEndPuncttrue
\mciteSetBstMidEndSepPunct{\mcitedefaultmidpunct}
{\mcitedefaultendpunct}{\mcitedefaultseppunct}\relax
\EndOfBibitem
\bibitem[Arimondo \emph{et~al.}(1977)Arimondo, Glorieux, and
  Oka]{arimondo:1977}
E.~Arimondo, P.~Glorieux, T.~Oka, \emph{Opt. Commun.} \textbf{1977}, \emph{23},
  369--372\relax
\mciteBstWouldAddEndPuncttrue
\mciteSetBstMidEndSepPunct{\mcitedefaultmidpunct}
{\mcitedefaultendpunct}{\mcitedefaultseppunct}\relax
\EndOfBibitem
\bibitem[Bauder \emph{et~al.}(1997)Bauder, Beil, Luckhaus, M{\"u}ller, and
  Quack]{bauder:1997}
A.~Bauder, A.~Beil, D.~Luckhaus, F.~M{\"u}ller, M.~Quack, \emph{J. Chem. Phys.}
  \textbf{1997}, \emph{106}, 7558--7570\relax
\mciteBstWouldAddEndPuncttrue
\mciteSetBstMidEndSepPunct{\mcitedefaultmidpunct}
{\mcitedefaultendpunct}{\mcitedefaultseppunct}\relax
\EndOfBibitem
\bibitem[Hollenstein \emph{et~al.}(1997)Hollenstein, Luckhaus, Pochert, Quack,
  and Seyfang]{hollenstein:1997}
H.~Hollenstein, D.~Luckhaus, J.~Pochert, M.~Quack, G.~Seyfang, \emph{Angew.
  Chem. Int. Ed.} \textbf{1997}, \emph{36}, 140--143\relax
\mciteBstWouldAddEndPuncttrue
\mciteSetBstMidEndSepPunct{\mcitedefaultmidpunct}
{\mcitedefaultendpunct}{\mcitedefaultseppunct}\relax
\EndOfBibitem
\bibitem[Costante \emph{et~al.}(1997)Costante, Hecht, Polavarapu, Collet, and
  Barron]{costante:1997}
J.~Costante, L.~Hecht, P.~L. Polavarapu, A.~Collet, L.~D. Barron, \emph{Angew.
  Chem. Int. Ed.} \textbf{1997}, \emph{36}, 885--887\relax
\mciteBstWouldAddEndPuncttrue
\mciteSetBstMidEndSepPunct{\mcitedefaultmidpunct}
{\mcitedefaultendpunct}{\mcitedefaultseppunct}\relax
\EndOfBibitem
\bibitem[Pitzer \emph{et~al.}(2013)Pitzer, Kunitski, Johnson, Jahnke, Sann,
  Sturm, Schmidt, Schmidt-B{\"o}cking, D{\"o}rner, Stohner, Kiedrowski,
  Reggelin, Marquardt, Schie{\ss}er, Berger, and Sch{\"o}ffler]{pitzer:2013}
M.~Pitzer, M.~Kunitski, A.~Johnson, T.~Jahnke, H.~Sann, F.~Sturm, L.~Schmidt,
  H.~Schmidt-B{\"o}cking, R.~D{\"o}rner, J.~Stohner, J.~Kiedrowski,
  M.~Reggelin, S.~Marquardt, A.~Schie{\ss}er, R.~Berger, M.~Sch{\"o}ffler,
  \emph{Science} \textbf{2013}, \emph{341}, 1096--1100\relax
\mciteBstWouldAddEndPuncttrue
\mciteSetBstMidEndSepPunct{\mcitedefaultmidpunct}
{\mcitedefaultendpunct}{\mcitedefaultseppunct}\relax
\EndOfBibitem
\bibitem[Pitzer \emph{et~al.}(2016)Pitzer, Kastirke, Kunitski, Jahnke, Bauer,
  Goihl, Trinter, Schober, Henrichs, Becht, Zeller, Gassert, Waitz, Kuhlins,
  Sann, Sturm, Wiegandt, Wallauer, Schmidt, Johnson, Mazenauer, Spenger,
  Marquardt, Marquardt, Schmidt-B{\"o}cking, Stohner, D{\"o}rner,
  Sch{\"o}ffler, and Berger]{pitzer:2015}
M.~Pitzer, G.~Kastirke, M.~Kunitski, T.~Jahnke, T.~Bauer, C.~Goihl, F.~Trinter,
  C.~Schober, K.~Henrichs, J.~Becht, S.~Zeller, H.~Gassert, M.~Waitz,
  A.~Kuhlins, H.~Sann, F.~Sturm, F.~Wiegandt, R.~Wallauer, L.~P.~H. Schmidt,
  A.~S. Johnson, M.~Mazenauer, B.~Spenger, S.~Marquardt, S.~Marquardt,
  H.~Schmidt-B{\"o}cking, J.~Stohner, R.~D{\"o}rner, M.~Sch{\"o}ffler,
  R.~Berger, \emph{ChemPhysChem.} \textbf{2016}, \emph{17}, 2465--2472\relax
\mciteBstWouldAddEndPuncttrue
\mciteSetBstMidEndSepPunct{\mcitedefaultmidpunct}
{\mcitedefaultendpunct}{\mcitedefaultseppunct}\relax
\EndOfBibitem
\bibitem[Daussy \emph{et~al.}(1999)Daussy, Marrel, Amy-Klein, Nguyen,
  Bord{\'e}, and Chardonnet]{daussy:1999}
C.~Daussy, T.~Marrel, A.~Amy-Klein, C.~T. Nguyen, C.~J. Bord{\'e},
  C.~Chardonnet, \emph{Phys. Rev. Lett.} \textbf{1999}, \emph{83},
  1554--1557\relax
\mciteBstWouldAddEndPuncttrue
\mciteSetBstMidEndSepPunct{\mcitedefaultmidpunct}
{\mcitedefaultendpunct}{\mcitedefaultseppunct}\relax
\EndOfBibitem
\bibitem[Ziskind \emph{et~al.}(2002)Ziskind, Daussy, Marrel, and
  Chardonnet]{ziskind:2002}
M.~Ziskind, C.~Daussy, T.~Marrel, C.~Chardonnet, \emph{Eur. Phys. J. D}
  \textbf{2002}, \emph{20}, 219--225\relax
\mciteBstWouldAddEndPuncttrue
\mciteSetBstMidEndSepPunct{\mcitedefaultmidpunct}
{\mcitedefaultendpunct}{\mcitedefaultseppunct}\relax
\EndOfBibitem
\bibitem[Quack and Stohner(2000)]{quack:2000}
M.~Quack, J.~Stohner, \emph{Phys. Rev. Lett.} \textbf{2000}, \emph{84},
  3807--3810\relax
\mciteBstWouldAddEndPuncttrue
\mciteSetBstMidEndSepPunct{\mcitedefaultmidpunct}
{\mcitedefaultendpunct}{\mcitedefaultseppunct}\relax
\EndOfBibitem
\bibitem[{J. K. Laerdahl and P. Schwerdtfeger and H. M.
  Quiney}(2000)]{laerdahl:2000a}
{J. K. Laerdahl and P. Schwerdtfeger and H. M. Quiney}, \emph{Phys. Rev. Lett}
  \textbf{2000}, \emph{84}, 3811--3814\relax
\mciteBstWouldAddEndPuncttrue
\mciteSetBstMidEndSepPunct{\mcitedefaultmidpunct}
{\mcitedefaultendpunct}{\mcitedefaultseppunct}\relax
\EndOfBibitem
\bibitem[Viglione \emph{et~al.}(2000)Viglione, Zanasi, Lazzeretti, and
  Ligabue]{viglione:2000}
R.~G. Viglione, R.~Zanasi, P.~Lazzeretti, A.~Ligabue, \emph{Phys. Rev. A}
  \textbf{2000}, \emph{62}, 052516\relax
\mciteBstWouldAddEndPuncttrue
\mciteSetBstMidEndSepPunct{\mcitedefaultmidpunct}
{\mcitedefaultendpunct}{\mcitedefaultseppunct}\relax
\EndOfBibitem
\bibitem[Quack and Willeke(2003)]{quack:2003}
M.~Quack, M.~Willeke, \emph{Helv. Chim. Acta} \textbf{2003}, \emph{86},
  1641--1652\relax
\mciteBstWouldAddEndPuncttrue
\mciteSetBstMidEndSepPunct{\mcitedefaultmidpunct}
{\mcitedefaultendpunct}{\mcitedefaultseppunct}\relax
\EndOfBibitem
\bibitem[Berger and Stuber(2007)]{berger:2007}
R.~Berger, J.~L. Stuber, \emph{Mol. Phys.} \textbf{2007}, \emph{105},
  41--49\relax
\mciteBstWouldAddEndPuncttrue
\mciteSetBstMidEndSepPunct{\mcitedefaultmidpunct}
{\mcitedefaultendpunct}{\mcitedefaultseppunct}\relax
\EndOfBibitem
\bibitem[Brueck \emph{et~al.}(2021)Brueck, Sahu, Gaul, and Berger]{bruck:2021}
S.~A. Brueck, N.~Sahu, K.~Gaul, R.~Berger, \emph{arXiv:2102.09897
  [physics.chem-ph]} \textbf{2021}\relax
\mciteBstWouldAddEndPuncttrue
\mciteSetBstMidEndSepPunct{\mcitedefaultmidpunct}
{\mcitedefaultendpunct}{\mcitedefaultseppunct}\relax
\EndOfBibitem
\bibitem[Rauhut and Schwerdtfeger(2021)]{rauhut:2021}
G.~Rauhut, P.~Schwerdtfeger, \emph{Phys. Rev. A} \textbf{2021}, \emph{103},
  042819\relax
\mciteBstWouldAddEndPuncttrue
\mciteSetBstMidEndSepPunct{\mcitedefaultmidpunct}
{\mcitedefaultendpunct}{\mcitedefaultseppunct}\relax
\EndOfBibitem
\bibitem[Hegstrom \emph{et~al.}(1980)Hegstrom, Rein, and
  Sandars]{hegstrom:1980}
R.~A. Hegstrom, D.~W. Rein, P.~G.~H. Sandars, \emph{J. Chem. Phys.}
  \textbf{1980}, \emph{73}, 2329--2341\relax
\mciteBstWouldAddEndPuncttrue
\mciteSetBstMidEndSepPunct{\mcitedefaultmidpunct}
{\mcitedefaultendpunct}{\mcitedefaultseppunct}\relax
\EndOfBibitem
\bibitem[Isaev \emph{et~al.}(2010)Isaev, Hoekstra, and Berger]{isaev:2010}
T.~A. Isaev, S.~Hoekstra, R.~Berger, \emph{Phys. Rev. A} \textbf{2010},
  \emph{82}, 052521\relax
\mciteBstWouldAddEndPuncttrue
\mciteSetBstMidEndSepPunct{\mcitedefaultmidpunct}
{\mcitedefaultendpunct}{\mcitedefaultseppunct}\relax
\EndOfBibitem
\bibitem[Garcia~Ruiz \emph{et~al.}(2020)Garcia~Ruiz, Berger, Billowes,
  Binnersley, Bissell, Breier, Brinson, Chrysalidis, Cocolios, Cooper,
  Flanagan, Giesen, de~Groote, Franchoo, Gustafsson, Isaev, {Koszor{\'u}s},
  Neyens, Perrett, Ricketts, Rothe, Schweikhard, Vernon, Wendt, Wienholtz,
  Wilkins, and Yang]{garciaruiz:2020}
R.~F. Garcia~Ruiz, R.~Berger, J.~Billowes, C.~L. Binnersley, M.~L. Bissell,
  A.~A. Breier, A.~J. Brinson, K.~Chrysalidis, T.~E. Cocolios, B.~S. Cooper,
  K.~T. Flanagan, T.~F. Giesen, R.~P. de~Groote, S.~Franchoo, F.~P. Gustafsson,
  T.~A. Isaev, {\'A}.~{Koszor{\'u}s}, G.~Neyens, H.~A. Perrett, C.~M. Ricketts,
  S.~Rothe, L.~Schweikhard, A.~R. Vernon, K.~D.~A. Wendt, F.~Wienholtz, S.~G.
  Wilkins, X.~F. Yang, \emph{Nature} \textbf{2020}, \emph{581}, 396--400\relax
\mciteBstWouldAddEndPuncttrue
\mciteSetBstMidEndSepPunct{\mcitedefaultmidpunct}
{\mcitedefaultendpunct}{\mcitedefaultseppunct}\relax
\EndOfBibitem
\bibitem[Schwerdtfeger \emph{et~al.}(2003)Schwerdtfeger, Gierlich, and
  Bollwein]{schwerdtfeger:2003}
P.~Schwerdtfeger, J.~Gierlich, T.~Bollwein, \emph{Angew. Chem. Int. Ed.}
  \textbf{2003}, \emph{42}, 1293--1296\relax
\mciteBstWouldAddEndPuncttrue
\mciteSetBstMidEndSepPunct{\mcitedefaultmidpunct}
{\mcitedefaultendpunct}{\mcitedefaultseppunct}\relax
\EndOfBibitem
\bibitem[Schwerdtfeger and Bast(2004)]{schwerdtfeger:2004b}
P.~Schwerdtfeger, R.~Bast, \emph{J. Am. Chem. Soc.} \textbf{2004}, \emph{126},
  1652--1653\relax
\mciteBstWouldAddEndPuncttrue
\mciteSetBstMidEndSepPunct{\mcitedefaultmidpunct}
{\mcitedefaultendpunct}{\mcitedefaultseppunct}\relax
\EndOfBibitem
\bibitem[Hirota(2012)]{hirota:2012}
E.~Hirota, \emph{Proc. Jpn. Acad. Ser. B} \textbf{2012}, \emph{88},
  120--128\relax
\mciteBstWouldAddEndPuncttrue
\mciteSetBstMidEndSepPunct{\mcitedefaultmidpunct}
{\mcitedefaultendpunct}{\mcitedefaultseppunct}\relax
\EndOfBibitem
\bibitem[Patterson \emph{et~al.}(2013)Patterson, Schnell, and
  Doyle]{patterson:2013}
D.~Patterson, M.~Schnell, J.~M. Doyle, \emph{Nature} \textbf{2013}, \emph{497},
  475--477\relax
\mciteBstWouldAddEndPuncttrue
\mciteSetBstMidEndSepPunct{\mcitedefaultmidpunct}
{\mcitedefaultendpunct}{\mcitedefaultseppunct}\relax
\EndOfBibitem
\bibitem[Medcraft \emph{et~al.}(2014)Medcraft, Wolf, and
  Schnell]{medcraft:2014}
C.~Medcraft, R.~Wolf, M.~Schnell, \emph{Angew. Chem. Int. Ed.} \textbf{2014},
  \emph{53}, 11656--11659\relax
\mciteBstWouldAddEndPuncttrue
\mciteSetBstMidEndSepPunct{\mcitedefaultmidpunct}
{\mcitedefaultendpunct}{\mcitedefaultseppunct}\relax
\EndOfBibitem
\bibitem[Schnell and K{\"u}pper(2011)]{schnell:2011}
M.~Schnell, J.~K{\"u}pper, \emph{Faraday Disc.} \textbf{2011}, \emph{150},
  33--49\relax
\mciteBstWouldAddEndPuncttrue
\mciteSetBstMidEndSepPunct{\mcitedefaultmidpunct}
{\mcitedefaultendpunct}{\mcitedefaultseppunct}\relax
\EndOfBibitem
\bibitem[Quack and Stohner(2000)]{quack:2000a}
M.~Quack, J.~Stohner, \emph{Z. Phys. Chem.} \textbf{2000}, \emph{214},
  675--703\relax
\mciteBstWouldAddEndPuncttrue
\mciteSetBstMidEndSepPunct{\mcitedefaultmidpunct}
{\mcitedefaultendpunct}{\mcitedefaultseppunct}\relax
\EndOfBibitem
\bibitem[Berger \emph{et~al.}(2001)Berger, Quack, and Stohner]{berger:2001}
R.~Berger, M.~Quack, J.~Stohner, \emph{Angew. Chem. Int. Ed.} \textbf{2001},
  \emph{40}, 1667--1670\relax
\mciteBstWouldAddEndPuncttrue
\mciteSetBstMidEndSepPunct{\mcitedefaultmidpunct}
{\mcitedefaultendpunct}{\mcitedefaultseppunct}\relax
\EndOfBibitem
\bibitem[Berger and Quack(2000)]{berger:2000a}
R.~Berger, M.~Quack, \emph{J. Chem. Phys.} \textbf{2000}, \emph{112},
  3148--3158\relax
\mciteBstWouldAddEndPuncttrue
\mciteSetBstMidEndSepPunct{\mcitedefaultmidpunct}
{\mcitedefaultendpunct}{\mcitedefaultseppunct}\relax
\EndOfBibitem
\bibitem[Gaul and Berger(2020)]{gaul:2020}
K.~Gaul, R.~Berger, \emph{J. Chem. Phys.} \textbf{2020}, \emph{152},
  044101\relax
\mciteBstWouldAddEndPuncttrue
\mciteSetBstMidEndSepPunct{\mcitedefaultmidpunct}
{\mcitedefaultendpunct}{\mcitedefaultseppunct}\relax
\EndOfBibitem
\bibitem[Gaul and Berger(2020)]{gaul:2020b}
K.~Gaul, R.~Berger, \emph{Mol. Phys.} \textbf{2020}, \emph{118}, e1797199\relax
\mciteBstWouldAddEndPuncttrue
\mciteSetBstMidEndSepPunct{\mcitedefaultmidpunct}
{\mcitedefaultendpunct}{\mcitedefaultseppunct}\relax
\EndOfBibitem
\bibitem[Autschbach \emph{et~al.}(2010)Autschbach, Zheng, and
  Schurko]{autschbach:2010}
J.~Autschbach, S.~Zheng, R.~W. Schurko, \emph{Concepts Magn. Reson. Part A
  Bridg. Educ. Res.} \textbf{2010}, \emph{36A}, 84--126\relax
\mciteBstWouldAddEndPuncttrue
\mciteSetBstMidEndSepPunct{\mcitedefaultmidpunct}
{\mcitedefaultendpunct}{\mcitedefaultseppunct}\relax
\EndOfBibitem
\bibitem[{Pyykk\"o}(2018)]{pyykko:2018}
P.~{Pyykk\"o}, \emph{Mol. Phys.} \textbf{2018}, \emph{116}, 1328--1338\relax
\mciteBstWouldAddEndPuncttrue
\mciteSetBstMidEndSepPunct{\mcitedefaultmidpunct}
{\mcitedefaultendpunct}{\mcitedefaultseppunct}\relax
\EndOfBibitem
\bibitem[Berger \emph{et~al.}(2005)Berger, Langermann, and van
  W{\"u}llen]{berger:2005}
R.~Berger, N.~Langermann, C.~van W{\"u}llen, \emph{Phys. Rev. A} \textbf{2005},
  \emph{71}, 042105\relax
\mciteBstWouldAddEndPuncttrue
\mciteSetBstMidEndSepPunct{\mcitedefaultmidpunct}
{\mcitedefaultendpunct}{\mcitedefaultseppunct}\relax
\EndOfBibitem
\bibitem[Berger and van W{\"u}llen(2005)]{berger:2005a}
R.~Berger, C.~van W{\"u}llen, \emph{J. Chem. Phys.} \textbf{2005}, \emph{122},
  134316\relax
\mciteBstWouldAddEndPuncttrue
\mciteSetBstMidEndSepPunct{\mcitedefaultmidpunct}
{\mcitedefaultendpunct}{\mcitedefaultseppunct}\relax
\EndOfBibitem
\bibitem[Nahrwold and Berger(2009)]{nahrwold:09}
S.~Nahrwold, R.~Berger, \emph{J. Chem. Phys.} \textbf{2009}, \emph{130},
  214101\relax
\mciteBstWouldAddEndPuncttrue
\mciteSetBstMidEndSepPunct{\mcitedefaultmidpunct}
{\mcitedefaultendpunct}{\mcitedefaultseppunct}\relax
\EndOfBibitem
\bibitem[Darquie \emph{et~al.}(2010)Darquie, Stoeffler, Shelkovnikov, Daussy,
  Amy-Klein, Chardonnet, Zrig, Guy, Crassous, Soulard, Asselin, Huet,
  Schwerdtfeger, Bast, and Saue]{darquie:2010}
B.~Darquie, C.~Stoeffler, A.~Shelkovnikov, C.~Daussy, A.~Amy-Klein,
  C.~Chardonnet, S.~Zrig, L.~Guy, J.~Crassous, P.~Soulard, P.~Asselin, T.~R.
  Huet, P.~Schwerdtfeger, R.~Bast, T.~Saue, \emph{{Chirality}} \textbf{2010},
  \emph{22}, 870--884\relax
\mciteBstWouldAddEndPuncttrue
\mciteSetBstMidEndSepPunct{\mcitedefaultmidpunct}
{\mcitedefaultendpunct}{\mcitedefaultseppunct}\relax
\EndOfBibitem
\bibitem[Cournol \emph{et~al.}(2019)Cournol, Manceau, Pierens, Lecordier, Tran,
  Santagata, Argence, Goncharov, Lopez, Abgrall, Coq, Targat, Martinez, Lee,
  Xu, Pottie, Hendricks, Wall, Bieniewska, Sauer, Tarbutt, Amy-Klein, Tokunaga,
  and Darqui{\'{e}}]{cournol:2019}
A.~Cournol, M.~Manceau, M.~Pierens, L.~Lecordier, D.~B.~A. Tran, R.~Santagata,
  B.~Argence, A.~Goncharov, O.~Lopez, M.~Abgrall, Y.~L. Coq, R.~L. Targat,
  H.~A. Martinez, W.~K. Lee, D.~Xu, P.~E. Pottie, R.~J. Hendricks, T.~E. Wall,
  J.~M. Bieniewska, B.~E. Sauer, M.~R. Tarbutt, A.~Amy-Klein, S.~K. Tokunaga,
  B.~Darqui{\'{e}}, \emph{Quantum Electron.} \textbf{2019}, \emph{49},
  288--292\relax
\mciteBstWouldAddEndPuncttrue
\mciteSetBstMidEndSepPunct{\mcitedefaultmidpunct}
{\mcitedefaultendpunct}{\mcitedefaultseppunct}\relax
\EndOfBibitem
\bibitem[Grabow \emph{et~al.}(2005)Grabow, Palmer, McCarthy, and
  Thaddeus]{grabow:2005}
J.-U. Grabow, E.~S. Palmer, M.~C. McCarthy, P.~Thaddeus, \emph{Rev. Sci.
  Instr.} \textbf{2005}, \emph{76}, 093106\relax
\mciteBstWouldAddEndPuncttrue
\mciteSetBstMidEndSepPunct{\mcitedefaultmidpunct}
{\mcitedefaultendpunct}{\mcitedefaultseppunct}\relax
\EndOfBibitem
\bibitem[van~de Meerakker \emph{et~al.}(2008)van~de Meerakker, Bethlem, and
  Meijer]{meerakker:2008}
S.~Y.~T. van~de Meerakker, H.~L. Bethlem, G.~Meijer, \emph{Nat. Phys.}
  \textbf{2008}, \emph{4}, 595--602\relax
\mciteBstWouldAddEndPuncttrue
\mciteSetBstMidEndSepPunct{\mcitedefaultmidpunct}
{\mcitedefaultendpunct}{\mcitedefaultseppunct}\relax
\EndOfBibitem
\bibitem[Quack \emph{et~al.}(2008)Quack, Stohner, and Willeke]{quack:2008}
M.~Quack, J.~Stohner, M.~Willeke, \emph{Annu. Rev. Phys. Chem.} \textbf{2008},
  \emph{59}, 741--769\relax
\mciteBstWouldAddEndPuncttrue
\mciteSetBstMidEndSepPunct{\mcitedefaultmidpunct}
{\mcitedefaultendpunct}{\mcitedefaultseppunct}\relax
\EndOfBibitem
\bibitem[Eibenberger \emph{et~al.}(2017)Eibenberger, Doyle, and
  Patterson]{eibenberger:2017}
S.~Eibenberger, J.~Doyle, D.~Patterson, \emph{Phys. Rev. Lett.} \textbf{2017},
  \emph{118}, 123002\relax
\mciteBstWouldAddEndPuncttrue
\mciteSetBstMidEndSepPunct{\mcitedefaultmidpunct}
{\mcitedefaultendpunct}{\mcitedefaultseppunct}\relax
\EndOfBibitem
\bibitem[Singh \emph{et~al.}(2023)Singh, Bergg{\"o}tz, Sun, and
  Schnell]{singh:2023}
H.~Singh, F.~Bergg{\"o}tz, W.~Sun, M.~Schnell, \emph{Angew. Chem. Int. Ed.}
  \textbf{2023},  e202219045\relax
\mciteBstWouldAddEndPuncttrue
\mciteSetBstMidEndSepPunct{\mcitedefaultmidpunct}
{\mcitedefaultendpunct}{\mcitedefaultseppunct}\relax
\EndOfBibitem
\bibitem[{de Nijs} \emph{et~al.}(2014){de Nijs}, Ubachs, and
  Bethlem]{nijs:2014}
A.~J. {de Nijs}, W.~Ubachs, H.~L. Bethlem, \emph{J. Mol. Spectrosc.}
  \textbf{2014}, \emph{300}, 79--85\relax
\mciteBstWouldAddEndPuncttrue
\mciteSetBstMidEndSepPunct{\mcitedefaultmidpunct}
{\mcitedefaultendpunct}{\mcitedefaultseppunct}\relax
\EndOfBibitem
\bibitem[Sahu \emph{et~al.}(2021)Sahu, Richardson, and Berger]{sahu:2021}
N.~Sahu, J.~O. Richardson, R.~Berger, \emph{J. Comput. Chem.} \textbf{2021},
  \emph{42}, 210--221\relax
\mciteBstWouldAddEndPuncttrue
\mciteSetBstMidEndSepPunct{\mcitedefaultmidpunct}
{\mcitedefaultendpunct}{\mcitedefaultseppunct}\relax
\EndOfBibitem
\bibitem[tur()]{turbomole7.5}
\emph{{TURBOMOLE V7.5 2020}, a development of {University of Karlsruhe} and
  {Forschungszentrum Karlsruhe GmbH}, 1989-2007, {TURBOMOLE GmbH}, since 2007;
  available from \\ {\tt https://www.turbomole.org}.}\relax
\mciteBstWouldAddEndPunctfalse
\mciteSetBstMidEndSepPunct{\mcitedefaultmidpunct}
{}{\mcitedefaultseppunct}\relax
\EndOfBibitem
\bibitem[van W{\"{u}}llen(2010)]{wullen:2010}
C.~van W{\"{u}}llen, \emph{Z. Phys. Chem} \textbf{2010}, \emph{224},
  413--426\relax
\mciteBstWouldAddEndPuncttrue
\mciteSetBstMidEndSepPunct{\mcitedefaultmidpunct}
{\mcitedefaultendpunct}{\mcitedefaultseppunct}\relax
\EndOfBibitem
\bibitem[Ahlrichs \emph{et~al.}(1989)Ahlrichs, B{\"a}r, H{\"a}ser, Horn, and
  K{\"o}lmel]{ahlrichs:1989}
R.~Ahlrichs, M.~B{\"a}r, M.~H{\"a}ser, H.~Horn, C.~K{\"o}lmel, \emph{Chem.
  Phys. Lett.} \textbf{1989}, \emph{162}, 165--169\relax
\mciteBstWouldAddEndPuncttrue
\mciteSetBstMidEndSepPunct{\mcitedefaultmidpunct}
{\mcitedefaultendpunct}{\mcitedefaultseppunct}\relax
\EndOfBibitem
\bibitem[Perdew \emph{et~al.}(1996)Perdew, Burke, and Ernzerhof]{perdew:1996}
J.~P. Perdew, K.~Burke, M.~Ernzerhof, \emph{Phys. Rev. Lett.} \textbf{1996},
  \emph{77}, 3865--3868\relax
\mciteBstWouldAddEndPuncttrue
\mciteSetBstMidEndSepPunct{\mcitedefaultmidpunct}
{\mcitedefaultendpunct}{\mcitedefaultseppunct}\relax
\EndOfBibitem
\bibitem[Adamo and Barone(1999)]{adamo:1999}
C.~Adamo, V.~Barone, \emph{J. Chem. Phys.} \textbf{1999}, \emph{110},
  6158--6170\relax
\mciteBstWouldAddEndPuncttrue
\mciteSetBstMidEndSepPunct{\mcitedefaultmidpunct}
{\mcitedefaultendpunct}{\mcitedefaultseppunct}\relax
\EndOfBibitem
\bibitem[Becke(1988)]{becke:1988}
A.~D. Becke, \emph{Phys. Rev. A} \textbf{1988}, \emph{38}, 3098--3100\relax
\mciteBstWouldAddEndPuncttrue
\mciteSetBstMidEndSepPunct{\mcitedefaultmidpunct}
{\mcitedefaultendpunct}{\mcitedefaultseppunct}\relax
\EndOfBibitem
\bibitem[Lee \emph{et~al.}(1988)Lee, Yang, and Parr]{lee:1988}
C.~Lee, W.~Yang, R.~G. Parr, \emph{Phys. Rev. B} \textbf{1988}, \emph{37},
  785--789\relax
\mciteBstWouldAddEndPuncttrue
\mciteSetBstMidEndSepPunct{\mcitedefaultmidpunct}
{\mcitedefaultendpunct}{\mcitedefaultseppunct}\relax
\EndOfBibitem
\bibitem[Vosko \emph{et~al.}(1980)Vosko, Wilk, and Nuisar]{vosko:1980}
S.~H. Vosko, L.~Wilk, M.~Nuisar, \emph{Can. J. Phys.} \textbf{1980}, \emph{58},
  1200--1211\relax
\mciteBstWouldAddEndPuncttrue
\mciteSetBstMidEndSepPunct{\mcitedefaultmidpunct}
{\mcitedefaultendpunct}{\mcitedefaultseppunct}\relax
\EndOfBibitem
\bibitem[Becke(1993)]{becke:1993}
A.~D. Becke, \emph{J. Chem. Phys.} \textbf{1993}, \emph{98}, 1372--1377\relax
\mciteBstWouldAddEndPuncttrue
\mciteSetBstMidEndSepPunct{\mcitedefaultmidpunct}
{\mcitedefaultendpunct}{\mcitedefaultseppunct}\relax
\EndOfBibitem
\bibitem[Sch{\"a}fer \emph{et~al.}(1994)Sch{\"a}fer, Huber, and
  Ahlrichs]{schafer:1994}
A.~Sch{\"a}fer, C.~Huber, R.~Ahlrichs, \emph{J. Chem. Phys.} \textbf{1994},
  \emph{100}, 5829--5835\relax
\mciteBstWouldAddEndPuncttrue
\mciteSetBstMidEndSepPunct{\mcitedefaultmidpunct}
{\mcitedefaultendpunct}{\mcitedefaultseppunct}\relax
\EndOfBibitem
\bibitem[Weigend \emph{et~al.}(1998)Weigend, H{\"a}ser, Patzelt, and
  Ahlrichs]{weigend:1998}
F.~Weigend, M.~H{\"a}ser, H.~Patzelt, R.~Ahlrichs, \emph{Chem. Phys. Lett.}
  \textbf{1998}, \emph{294}, 143--152\relax
\mciteBstWouldAddEndPuncttrue
\mciteSetBstMidEndSepPunct{\mcitedefaultmidpunct}
{\mcitedefaultendpunct}{\mcitedefaultseppunct}\relax
\EndOfBibitem
\bibitem[Figgen \emph{et~al.}(2009)Figgen, Peterson, Dolg, and
  Stoll]{figgen:2009}
D.~Figgen, K.~A. Peterson, M.~Dolg, H.~Stoll, \emph{J. Chem. Phys.}
  \textbf{2009}, \emph{130}, 164108\relax
\mciteBstWouldAddEndPuncttrue
\mciteSetBstMidEndSepPunct{\mcitedefaultmidpunct}
{\mcitedefaultendpunct}{\mcitedefaultseppunct}\relax
\EndOfBibitem
\bibitem[Pollak and Weigend(2017)]{pollak:2017}
P.~Pollak, F.~Weigend, \emph{J. Chem. Theory Comput.} \textbf{2017}, \emph{13},
  3696--3705\relax
\mciteBstWouldAddEndPuncttrue
\mciteSetBstMidEndSepPunct{\mcitedefaultmidpunct}
{\mcitedefaultendpunct}{\mcitedefaultseppunct}\relax
\EndOfBibitem
\bibitem[Slater(1951)]{slater:1951}
J.~C. Slater, \emph{Phys. Rev.} \textbf{1951}, \emph{81}, 385--390\relax
\mciteBstWouldAddEndPuncttrue
\mciteSetBstMidEndSepPunct{\mcitedefaultmidpunct}
{\mcitedefaultendpunct}{\mcitedefaultseppunct}\relax
\EndOfBibitem
\bibitem[Visscher and Dyall(1997)]{visscher:1997}
L.~Visscher, K.~G. Dyall, \emph{At. Data Nucl. Data Tables} \textbf{1997},
  \emph{67}, 207--224\relax
\mciteBstWouldAddEndPuncttrue
\mciteSetBstMidEndSepPunct{\mcitedefaultmidpunct}
{\mcitedefaultendpunct}{\mcitedefaultseppunct}\relax
\EndOfBibitem
\end{mcitethebibliography}
\end{document}